\documentstyle[12pt,epsf]{article}
\newcommand\pubnumber{SLAC-PUB-8351}
\newcommand\pubdate{February, 2000}

\def\Title#1{\begin{center} {\Large #1 } \end{center}}
\def\Author#1{\begin{center}{ \sc #1} \end{center}}
\def\Address#1{\begin{center}{ \it #1} \end{center}}

\def\doeack{\footnote{Work supported by the Department of Energy,
                     contract DE--AC03--76SF00515.}}
\def\SLAC{Stanford Linear Accelerator Center\\
    Stanford University, Stanford, California 94309 USA}
\newcommand\pubblock{\rightline{\begin{tabular}{l} \pubnumber\\
         \pubdate  \end{tabular}}}
\newenvironment{Abstract}{\begin{quotation} \begin{center}
                       ABSTRACT
     \end{center}\bigskip  }{\end{quotation}}

\def\beq{\begin{equation}}
\def\eeq#1{\label{#1}\end{equation}}
\def\eeqn{\end{equation}}
\def\beqa{\begin{eqnarray}}
\def\eeqa#1{\label{#1}\end{eqnarray}}
\def\eeqan{\end{eqnarray}}
\def\CR{\nonumber \\ }
\def\leqn#1{(\ref{#1})}
\let\bar=\overbar
\def\Dslash{\not{\hbox{\kern-4pt $D$}}}
\def\dslash{\not{\hbox{\kern-2pt $\del$}}}
\def\half{\frac{1}{2}}

\def\O{{\cal O}}

\def\del{\partial}

\def\Pl{{\mbox{\scriptsize Pl}}}
\def\GUT{{\mbox{\scriptsize GUT}}}

\def\ee{e^+e^-}
\def\sstw{\sin^2\theta_w}

\def\mz{m_Z}

\def\mw{m_W}

\def\msbar{\bar{M \kern -2pt S}}
\def\msb{{\bar{\ssstyle M \kern -1pt S}}}

\def\s#1{\widetilde{#1}}
\def\etal{{\it et al.}}

\begin{document}
\begin{titlepage}
\pubblock

\vfill
\Title{Theoretical Summary Lecture for EPS HEP99}
\vfill
\Author{Michael E. Peskin\doeack}
\Address{\SLAC}
\vfill
\begin{Abstract}
This is the proceedings article for the concluding lecture of the 1999 High 
Energy Physics Conference of the European Physical Society.  In this article,
I review a number of topics that were highlighted at the meeting and have
more general importance in high energy physics.  The major topics discussed
are (1)~precision electroweak physics, (2) CP violation, (3) new directions 
in QCD, (4)~supersymmetry spectroscopy, and (5) the experimental physics of 
extra dimensions.
\end{Abstract}
\vfill
\begin{center}
invited lecture presented at the \\
International Europhysics Conference on High-Energy Physics\\
July 15-21, 1999, Tampere, Finland
\end{center}

\vfill
\end{titlepage}
\def\thefootnote{\fnsymbol{footnote}}
\setcounter{footnote}{0}

\section{Introduction}

In this theoretical 
summary lecture at the High Energy Physics 99 conference of the 
European Physical Society, I am charged to review some of the 
new conceptual developments
presented at this conference.  At the same time, I would like to review
more generally the progress of high-energy physics over the past year, and
to highlight areas in which our basic understanding has been affected by 
the new developments.  There is no space here for a status report on the 
whole field.  But I would like to give extended discussion to five areas
that I think have special importance this year.  These are (1) precision
electroweak physics, which celebrates its tenth anniversary this summer;
(2) CP violation, which entered a new era this summer with the inauguration
of the SLAC and KEK B-factories; (3) QCD, which now branches into new lines
of investigation; and two rapidly developing topics from physics beyond 
the Standard Model, (4) supersymmetry spectroscopy and (5) the experimental
study of extra dimensions.  A sixth important topic, that of neutrino
masses and mixing, is covered in the experimental summary talk of 
Lorenzo Foa \cite{Foa}. The location of the conference in Finland makes it
appropriate to end the lecture with a Lutheran sermon.

\section{Precision Electroweak Physics}

This summer marks the tenth anniversary of a watershed in high-energy
physics that took place in the summer and fall of 1989.  In that year,
the UA2 and CDF experiments announced the first truly precision measurements
of the $W$ boson mass.  These experiments  also pushed the mass of the top 
quark above 60 GeV, thus insuring that radiative corrections due to the top
quark would play an important role in the  interpretation of  weak interaction
measurements.  SLC and LEP began their high-statistics study of $Z^0$ 
resonance in $\ee$ annihilation.  The data from these machines rapidly 
produced a $Z^0$ mass accurate to four significant figures, limited the 
number of light neutrinos to 3, and began the program of precisely testing 
the weak-interaction couplings.  In the spring of 1989, it was permissible to 
believe that the $W$ and $Z$ bosons were composite and that the gauge 
symmetry of the weak interactions was merely a low-energy approximation.
Today, because of the experimental program set in motion that year,
this is no longer an option.

\begin{figure}
\begin{center}
\leavevmode
{\epsfxsize=5.00truein \epsfbox{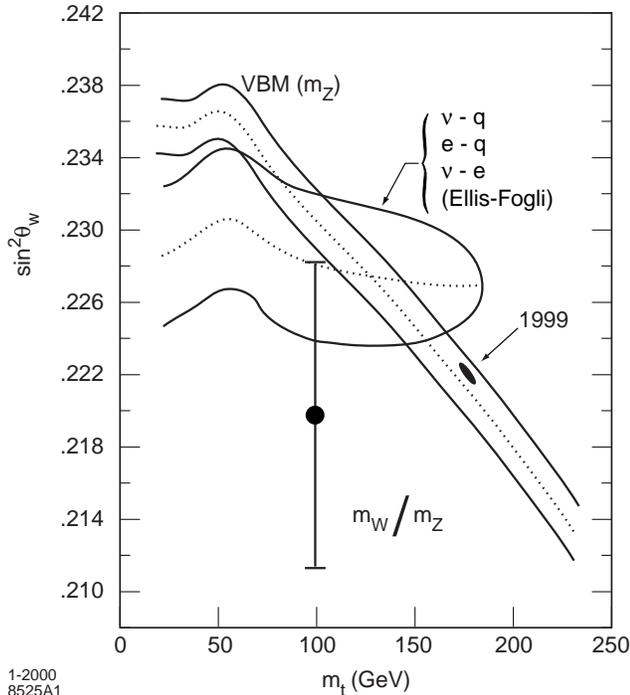}}
\end{center}
\caption{Constraints on $\sstw$ as a function of the top quark mass,
        shown by Altarelli at the 1989 Lepton-Photon conference [2].
        The small dot marked 1999 shows our current knowledge.}
\label{fig:Altar}
\end{figure}

It is interesting to contrast our knowledge of the weak interaction parameters
in 1989 with our knowledge today.  In Figure~\ref{fig:Altar}, I show the 
summary of the constraints on the weak interaction mixing angle $\sstw$
that were presented by Altarelli at the 1989 Lepton-Photon 
conference \cite{Altarelli}.
The vertical axis shows $\sstw$ in Sirlin's definition \cite{Sirlin}
\beq
     \sstw\bigr|_{{\mbox{\scriptsize Sirlin}}} = 1 - {\mw^2\over \mz^2} \ .
\eeq{sirlin}
The horizontal axis shows the top quark mass, which enters the comparison
of weak interaction processes through $\O(\alpha)$ vacuum polarization 
diagrams.  The constraints shown were all new:
the Ellis-Fogli fit to deep-inelastic neutrino and electron 
scattering \cite{EF}, the UA2 and CDF measurements of 
$\mw/\mz$ \cite{UA2,CDFW}, and the
SLC measurement of the $Z^0$ mass \cite{SLCZ}.
 Contrast the precision of this figure,
remarkable at the time, with the small trace labeled `1999'. This dot
represents our current knowledge of $m_t$ and $\sstw$.

%%%%%%%%%%%%%%%%%%%%%%%%%%%%%%%%%%%%%%%%%%%%%%%%%%%%%%%%%%%%%%%%%%%%%%%
%%%%%%%%%%%%%%%%%%%%%%%%%%%%%%%%%%%%%%%%%%%%%%%%%%%%%%%%%%%%%%%%%%%%%%%%%%%

Within a month after Altarelli showed this figure, the LEP collider began
its high-statistics study of the $Z^0$ resonance.  The precision of these
experiments, and their remarkable agreement with the Standard Model 
predictions, has led to a major change in the way that we think about the
weak interactions.  Today, we regard $\mz$ as a fundamental constant of Nature,
determined to precision of five significant figures and thus standing on a 
par with $\alpha$ and $G_F$.  The precise values of these three parameters
fix the tree-level predictions of the Standard Model. Experiments can then
focus on possible small deviations from these predictions, which might
be due to the radiative corrections of the Standard Model, or to new physics.

There are several schemes for presenting relatively model-independent 
constraints on weak interaction radiative corrections.  My favorite is to 
parametrize the $W$ and $Z$ boson vacuum polarization diagrams in terms of
two parameters $S$ and $T$, scaled so that an effect of order 1 in these 
parameters corresponds to a correction of order $\alpha$ in the electroweak
observables \cite{PT}.  (Other similar parameters sets are defined in
\cite{LK,AB}.)
The parameter $S$ is weak-isospin conserving and 
measures the overall size of a new physics sector; the parameter $T$ measures
the extent of its weak-isospin violation.  The zero point of $S$ and $T$ is 
fixed by convention; for my discussion here, I will fix it to correspond to the
minimal Standard Model with a top quark mass of 175 GeV and a Higgs boson 
mass of 100 GeV.

%%%%%%%%%%%%%%%%%%%%%%%%%%%%%%%%%%%%%%%%%%%%%%%%%%%%%%%%%%%%%%%%%%%%%%%
\begin{figure*}
\begin{center}
\leavevmode
{\epsfxsize=4.50truein \epsfbox{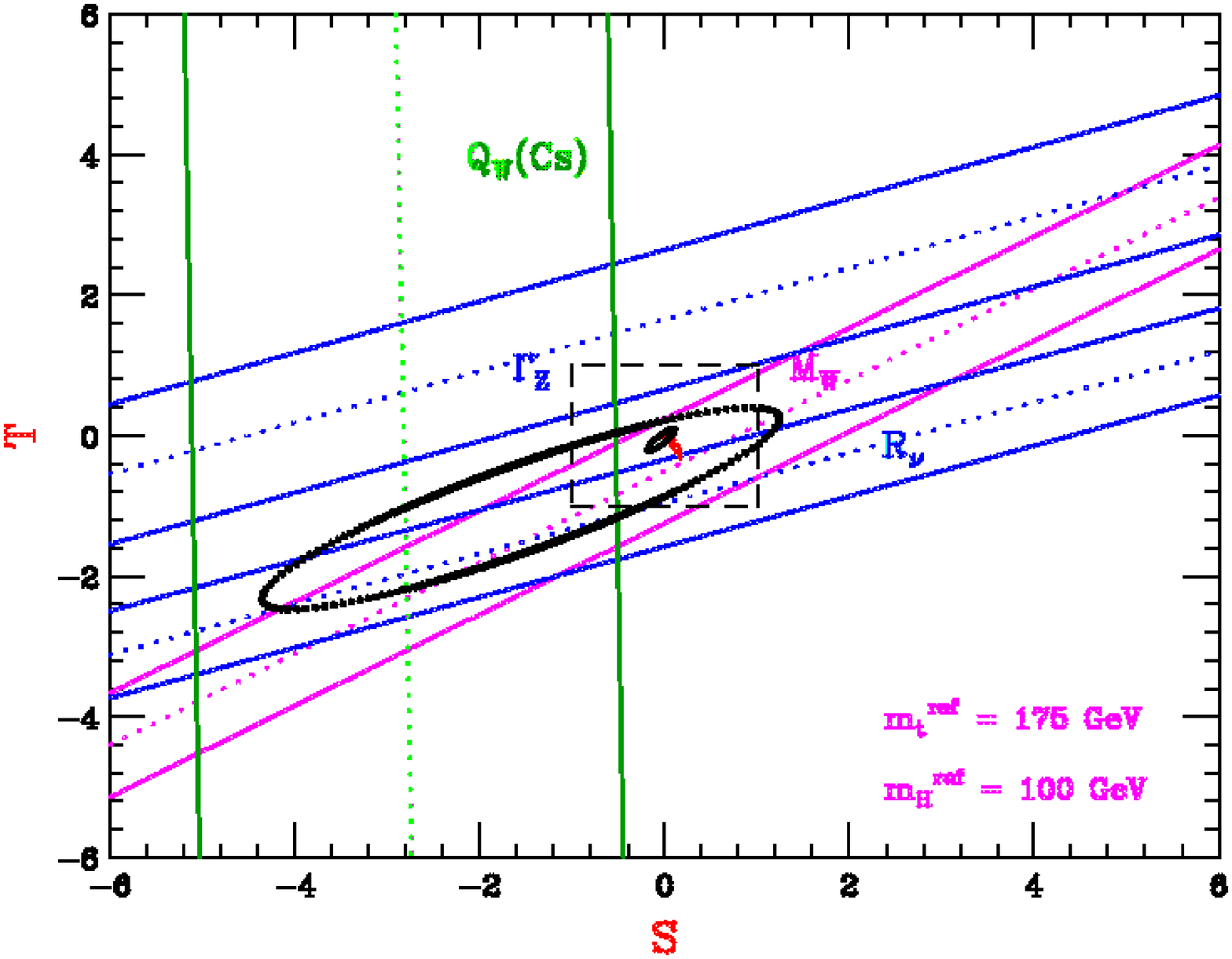}}
{\epsfxsize=4.50truein \epsfbox{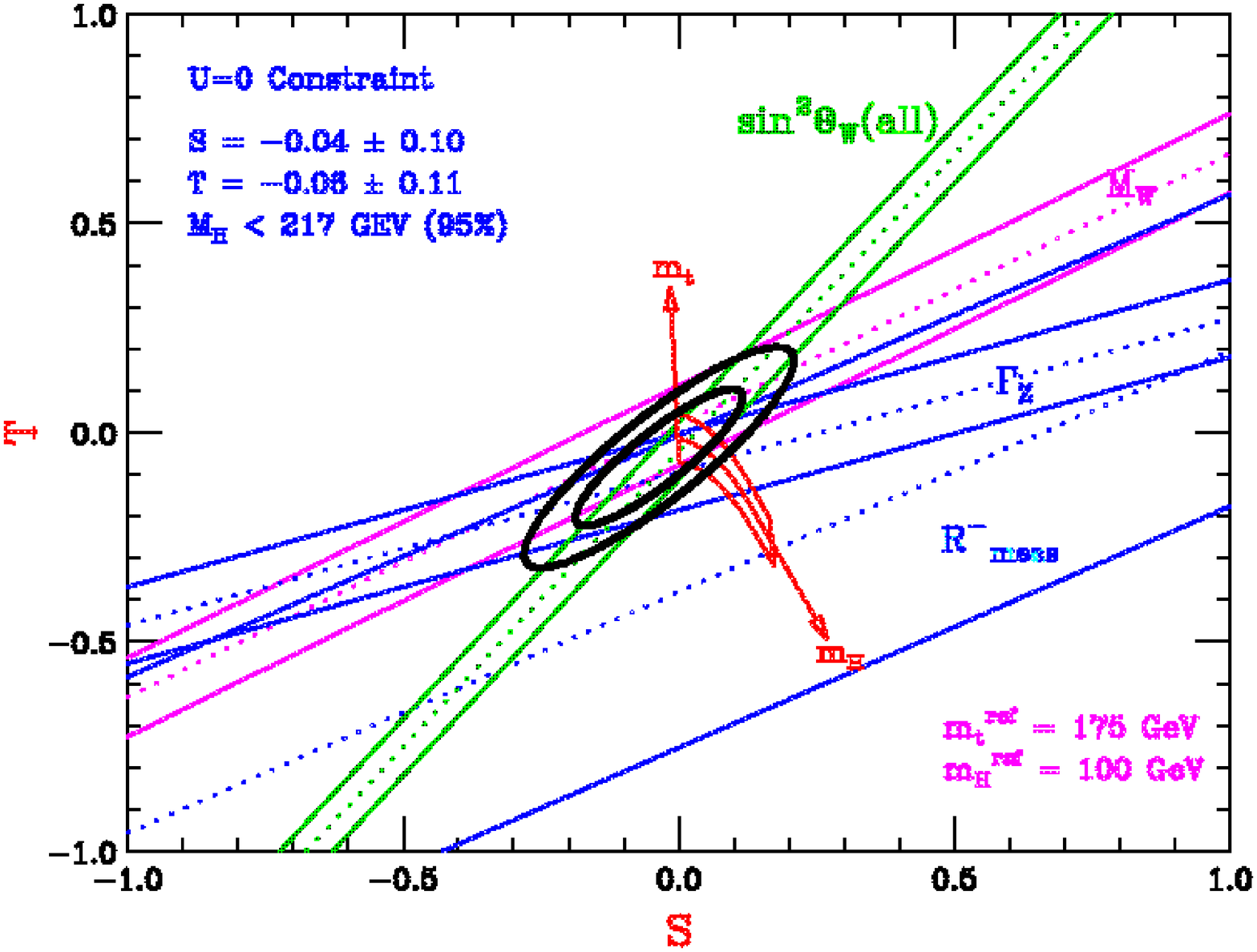}}
\end{center}
\caption{Fits of the corpus of precision electroweak data to the parameters
  $S$ and $T$, for the data available in the summer of 1989 and the summer
  of 1999 [11].}
\label{fig:Swartz}
\end{figure*}
%%%%%%%%%%%%%%%%%%%%%%%%%%%%%%%%%%%%%%%%%%%%%%%%%%%%%%%%%%%%%%%%%%%%%%%%%%
%
In Figure~\ref{fig:Swartz}, I present the $S$, $T$ fit to the corpus of 
precision 
electroweak data from the summer of 1989 and from the summer of 1999, both
prepared by Morris Swartz \cite{Swartz}.  The bottom figure fits into 
the small dashed rectangle in the top figure.   The current best values
are
\beq
      S = -0.04 \pm 0.10    \  , \qquad   T =  -0.06 \pm  0.11 \ . 
\eeq{STvals}

What have we learned from this dramatic improvement in our experimental 
knowledge?  I extract three morals:

{\bf First}, we have learned that
the predictions of the minimal Standard Model are amazingly 
successful!  I remind you that the agreement of predictions to better than
order-1 on the $S,T$ plot requires radiative corrections, and that the
experimental success thus tests the Standard Model at the loop level. 
This aspect is discussed in more detail in \cite{SirlinLP99}.   

This level of agreement causes deep difficulty for many schemes of 
physics beyond the Standard Model.  Certain models of new physics 
have the property that they `decouple' when the scale of new physics becomes
large.  In brief, this means that new particles of mass $M$ produce
 corrections to $S$ and $T$ that are
of order
\beq
          S\ ,\ T \sim {1\over 4\pi} {\mz^2\over M^2} \ 
\eeq{STdecoup}
for $M \gg \mz$.
The precision electroweak results imply that, generically, any model of 
new physics that does not naturally decouple in this way is excluded.
This is a severe setback for technicolor models, models with a fourth 
generation of quarks and leptons, and models in which quarks and leptons
are composite.  Models with decoupling are typically also models in which
the Higgs boson is a fundamental weakly-coupled scalar particle, so 
the precision electroweak results support this 
hypothesis.

There are still some notable discrepancies in the picture.  In his review
at HEP99,
Mnich presented a combined value of the $Z$ polarization asymmetry of $b$
quarks \cite{Mnich},
\beq
     A_b = 0.893 \pm 0.016 \ \qquad (SM: \ 0.935) \ .
\eeq{Abval}
On the other hand, the value of the $b$ fraction of hadronic $Z$ decays
has now settled down to 
\beq
    R_b = 0.21642 \pm 0.00073 \ ,
\eeq{Rbval}
which agrees with the standard model to 0.3\% accuracy.  For comparison,
technicolor models typically predict a 3\% discrepancy \cite{Chivuk}.
A theorist who wanted to pursue this matter could construct a model with a
large deviation in $A_b$ and no deviation in $R_b$, but essentially all
models constructed in advance of the data predicted the opposite pattern.

Further improvements in the precision of the comparison of electroweak data
to the Standard Model will require a more precise determination of the 
renormalization of $\alpha$ from $Q^2 = 0$ to $Q^2 = \mz^2$.  This requires
a precise knowledge of the total cross section for $\ee$ annihilation to
hadrons through this energy range.  We are fortunate that the Beijing Electron
Synchrotron has made this measurement a focus of its experimental program
and expects to dramatically improve our knowledge of the cross section in 
the charm threshold region  \cite{Zhu}.

{\bf Second},  we have determined 
the parameters of the Standard Model to
remarkable precision.  In particular, we now have accurate values for the 
fundamental Standard Model coupling constants.  In terms of $\msbar$ 
couplings at the scale $\mz$,
\beqa
         \alpha' &=&  1/ 98.42 \pm 0.27 \CR
         \alpha_w &=&  1/ 29.60 \pm 0.08 \CR
         \alpha_s &=&  1/ 8.40 \pm 0.14      \ .
\eeqa{SMcouplings}

At the same time, the precision electroweak data constraints give us 
information on the mass of the Higgs boson.  In a fit to the minimal
Standard Model, one now finds \cite{Mnich}
\beq
         m_H < 245 \ \mbox{GeV} \ \qquad \mbox{(95\%\ conf.)} 
\eeq{mHval}
These results generalize to models with multiple Higgs bosons.  Assume,
for example, that there are several Higgs bosons $\phi_i$ with vacuum
expectation values $v_i$, and set
\beq
              v = 246 \ \mbox{GeV} \ .
\eeq{vval}
Then there is a sum rule $\sum_i v_i^2 = v^2$~\cite{GH}, and the precision
electroweak data adds the information
\beq
               \sum_i  \left({v_i^2\over v^2}\right) \log 
                            {m_i \over \mbox(245 GeV)} <  0  \ .
\eeq{massHsum}
In principle, new interactions at high energy can contribute to the right-hand
side of \leqn{massHsum}; however, in the simplest models, these contributions
are negative (corresponding to positive contributions to $S$) \cite{PT,AT}.

Both of the new pieces of information are encouraging for supersymmetry.
The precisely known values of the couplings are consistent with a grand
unification of couplings if the renormalization group equations of 
supersymmetry are used in the comparison \cite{sRGE}. 
 Supersymmetric grand unified
theories require a Higgs boson below 180 GeV\cite{Gordy,X,Y}.  
A useful reference value is the prediction of the minimal supersymmetric
generalization of the Standard Model (MSSM), for large superparticle masses
and reasonably large $\tan\beta$,  $m_h \sim 120$ GeV.  This should be 
compared to the new direct search limit on the Higgs boson mass announced at 
HEP99 \cite{Higgs}
\beq
             m_h > 95.2 \ \mbox{GeV} \ \qquad \mbox{(95\%\ conf.)} \ .
\eeq{Higgslimit}

\begin{figure}
\begin{center}
\leavevmode
{\epsfxsize=5.50truein \epsfbox{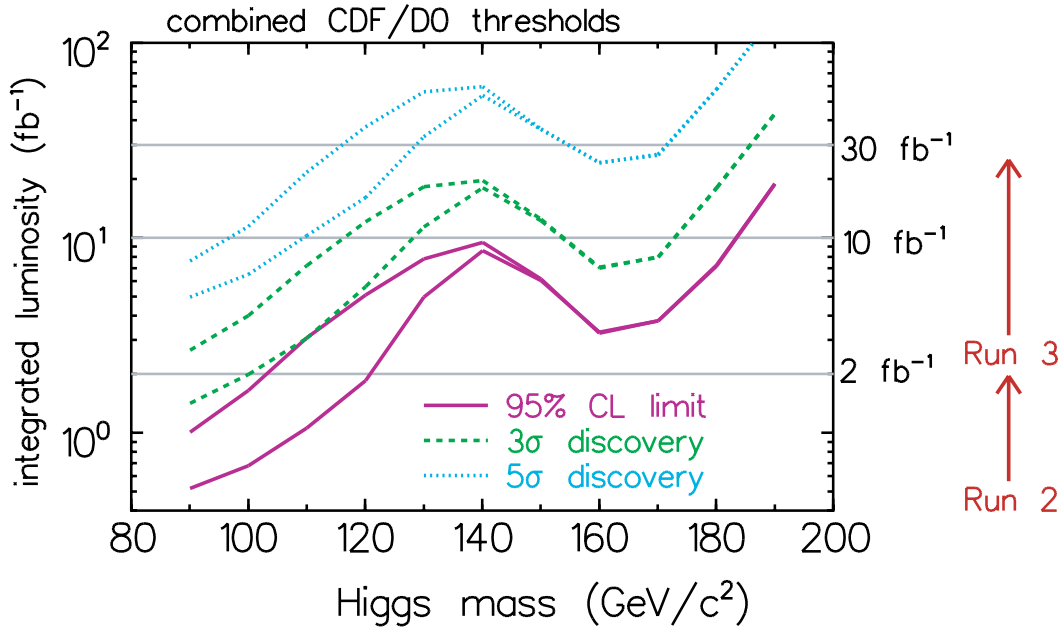}}
\end{center}
\caption{Expected capability of the Run II Tevatron experiments to observe
the minimal Standard Model Higgs boson, at various levels of integrated 
luminosity [23].  The lower curve shows a preliminary improved analysis
based on neural network techniques.}
\label{fig:TevHiggs}
\end{figure}

I am still hoping that the Higgs boson will be found at LEP before its time
runs out.  If not, there is a new entrant into the race to discover the Higgs
boson, the Run II Tevatron experiments.  A new analysis of Tevatron 
capabilities takes account of many possible improvements from earlier
studies \cite{Roco}.
  The Higgs is searched for in the decay mode $h^0 \to b\bar b$
in the reaction $p\bar p \to W h$, with a leptonic decay of the $W$, and
 $p\bar p \to Z h$, with a $\nu\bar \nu$ decay of the $Z$.  The expected 
improvements in vertex identification and $b\bar b$ mass resolution are 
included, and the new ability to trigger on displayed vertices plays an 
important role.  The expected sensitivity of the Tevatron experiments
is shown in Figure~\ref{fig:TevHiggs}.  If the Tevatron can accumulate
20 fb$^{-1}$ of integrated luminosity, it should be able to find the Higgs
boson in the whole region expected in the MSSM, and in most of the region
expected for any model with a weakly-coupled Higgs boson.

%%%%%%%%%%%%%%%%%%%%%%%%%%%%%%%%%%%%%%%%%%%%%%%%%%%%%%%%%%%%%%%%%%%%%%%
%%%%%%%%%%%%%%%%%%%%%%%%%%%%%%%%%%%%%%%%%%%%%%%%%%%%%%%%%%%%%%%%%%%%%%%%%%%

\begin{figure*}
\begin{center}
\leavevmode
{\epsfxsize=2.90truein \epsfbox{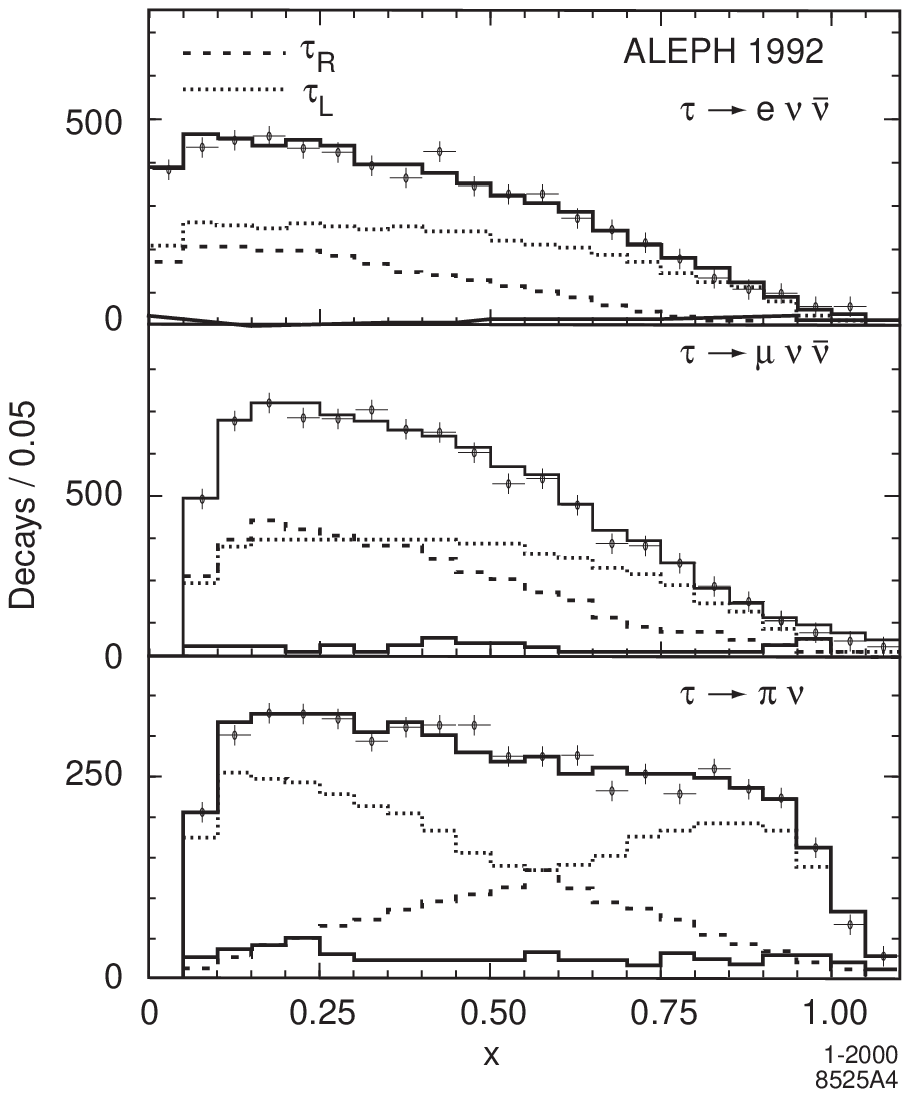}}\hspace*{0.0in}
{\epsfxsize=3.20truein \epsfbox{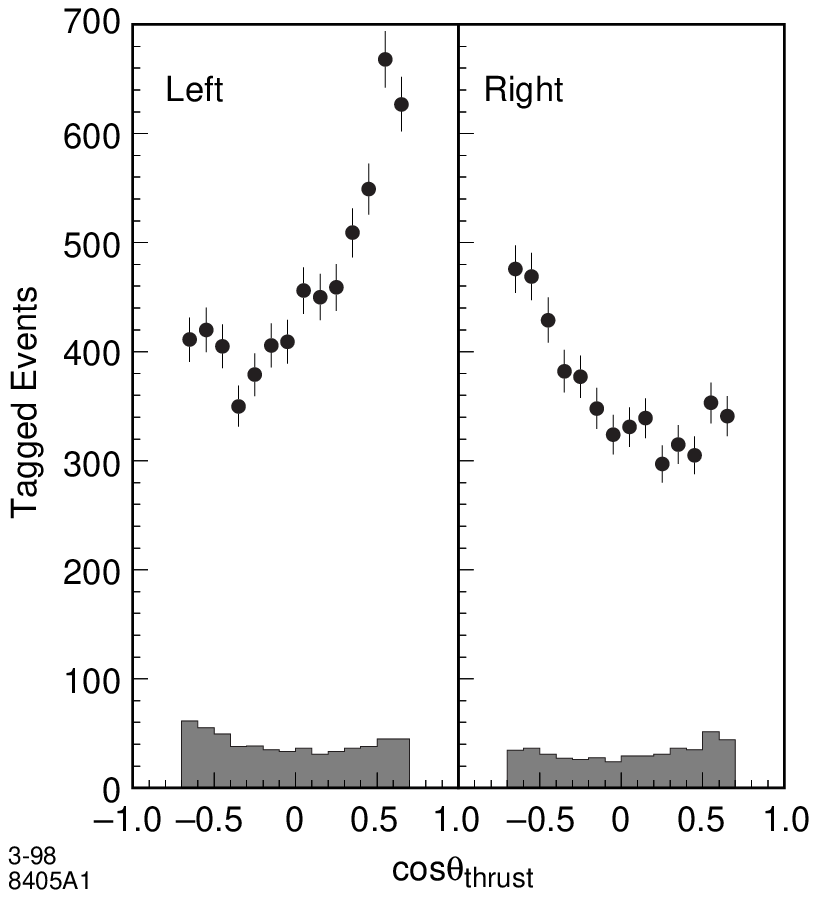}}
{\epsfxsize=2.60truein \epsfbox{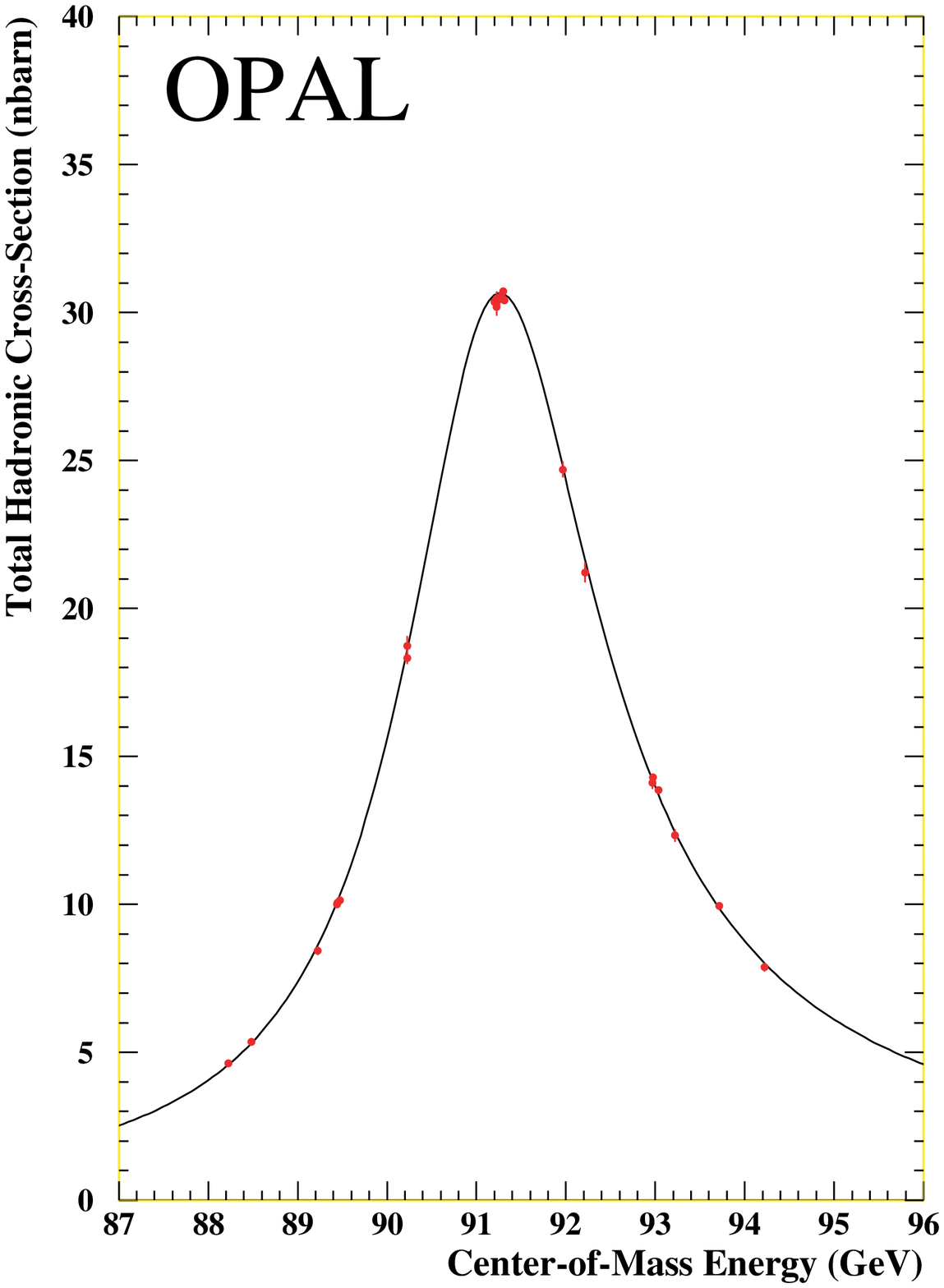}}\hspace*{0.6in}
{\epsfxsize=3.30truein \epsfbox{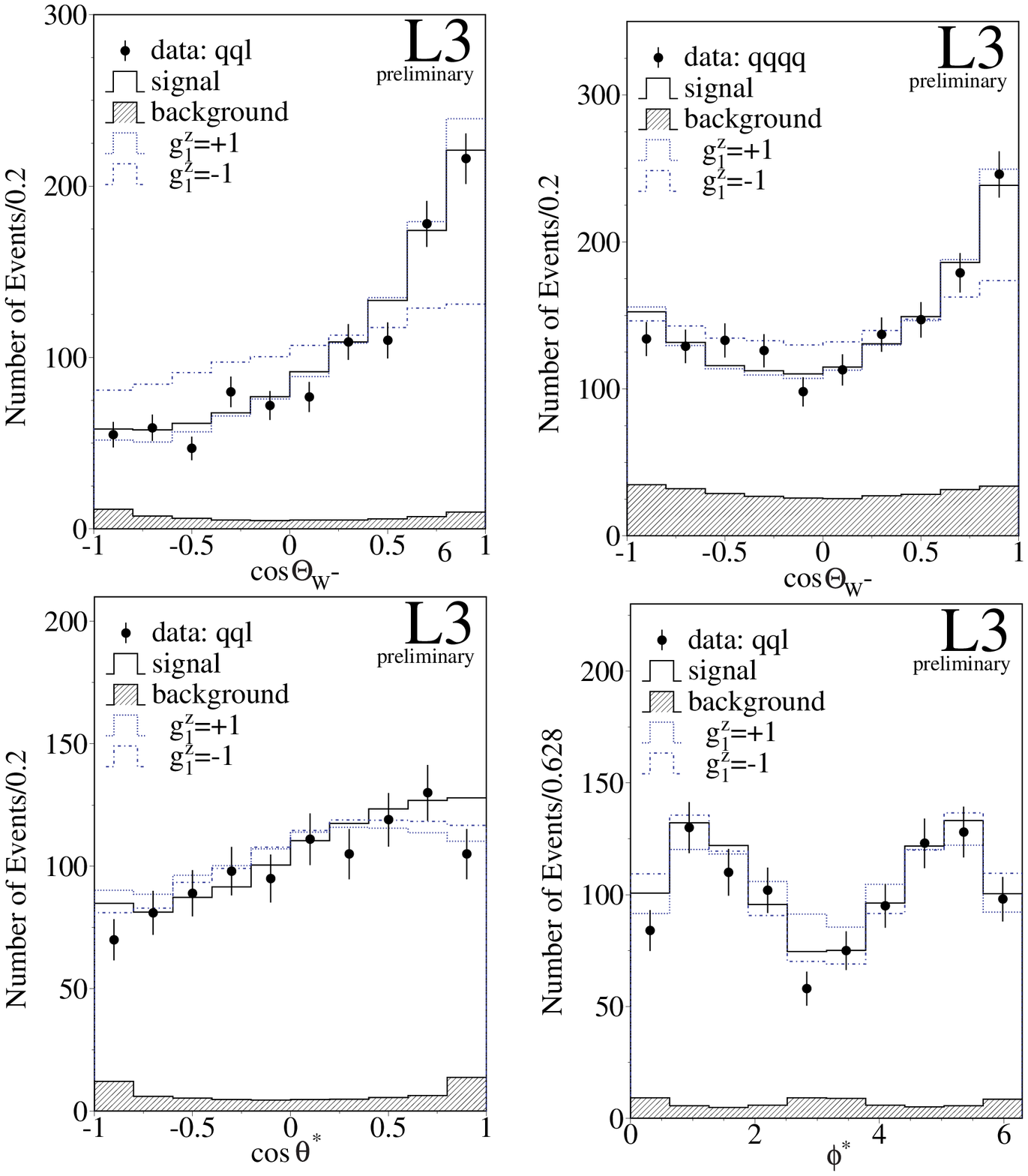}}
\end{center}
\caption{Four figures which display how the Standard Model works at the
$Z^0$ resonance and at higher energy.  See the text for more details.}
\label{fig:montage1}
\end{figure*}

{\bf Third}, we have acquired a tactile appreciation for the Standard
Model couplings to quarks and leptons.  The beautiful experiments at the 
$Z^0$ resonance do not simply give the $Z^0$ couplings as outputs of a fit;
they show directly how the Standard Model works.  From the many remarkable
plots that have come out of the LEP and SLC program, I show three of my 
favorites in Figure~\ref{fig:montage1}.  The upper left
shows the ALEPH determination of the $\tau$ polarization at the $Z^0$, in which
three decay modes, each with its own characteristic physics, show a 14\%
excess of $\tau_L$ over $\tau_R$.   The upper right shows the
profound effect of electron beam polarization on the $b$ angular distribution,
characteristic of the almost complete dominance of $b_L$ in $Z^0$ decays, as
observed by the SLD experiment.
The lower left  shows the OPAL determination of the resonance
line-shape for $Z^0$ decays to hadrons.  The remarkable agreement shown 
reflects our understanding of all three of the fundamental interactions,
weak, through the gross form and precision radiative corrections, strong,
through the order $\alpha_s$ correction to the decay width to quarks, and
electromagnetic, through the distortion of the line-shape by initial-state
radiation.  In the lower right, I add a new figure shown for 
the first time at HEP99, the measurement of the $W$ boson production and
decay angular distributions by L3~\cite{Macchiolo}.
This shows the forward peak in the production angle 
expected from neutrino exchange, and the correct proportion
 of events with central values of the decay angle, characteristic of
longitudinal $W$ polarization.  All four plots speak directly to the 
basic underlying physics.  It is no longer a tenable position
that the Standard Model is a `social construct'; we see its reality before
our eyes.

%%%%%%%%%%%%%%%%%%%%%%%%%%%%%%%%%%%%%%%%%%%%%%%%%%%%%%%%%%%%%%%%%%%%%%%
%%%%%%%%%%%%%%%%%%%%%%%%%%%%%%%%%%%%%%%%%%%%%%%%%%%%%%%%%%%%%%%%%%%%%%%%%%%

\section{CP violation}

Just as this year marks the completion of an era in electroweak physics,
it marks the beginning of an era in the study of CP violation.  We have seen
one of the major questions about CP violation finally answered, and we have
seen the first results from new facilities that will dramatically reshape our
experimental knowledge.  

To put both developments in perspective, I will 
begin my discussion with a capsule history of CP violation.  The phenomenon
was discovered in 1964, in the classic experiment of Christensen, Cronin,
Fitch, and Turlay \cite{CCFT}.  Almost immediately thereafter, Wolfenstein
asked a crucial question \cite{Wolf}:  Is CP violation a part of 
the weak interactions,
or is it due to a new interaction at very small distances?  Over the years,
many models have been proposed in which CP violation arises from 
weak-interaction couplings of particles with masses of the order of $\mw$;
the Kobayashi-Maskawa model \cite{KM},
 in which CP violation is due to quark mixing,
the Weinberg model \cite{WeinbergH}, in which CP violation is due 
to Higgs boson mixing, and
other models in which CP violation comes from phases in the mixing of more
exotic species.  Behind all of these, though, lurked the possibility of a 
`superweak' origin for CP violation, in which CP violation arose from a new
hard coupling which affected only the $K^0$--$\bar{K}{}^0$ mixing.

In 1979, Gilman and Wise proposed a crucial test of the weak-interaction origin
of CP violation \cite{GW}.
  They showed that such theories typically predict a small
but nonzero influence of CP violation on the $K^0$ decay amplitudes through
 the parameter $\epsilon'$.  In 1988, the CERN NA31 experiment found a nonzero
value for $\epsilon'$ \cite{NA31}, but this result was not confirmed 
by the competing
experiment E731 at Fermilab \cite{firstWinstein}.  
Finally, this year, the new high precision experiments
NA48 and KTeV agree that $\epsilon'$ is nonzero and find quite similar 
values \cite{NA48,KTeV}.
The new world average presented at HEP99 is \cite{Buchalla}
\beq
      \epsilon'/\epsilon = ( 21.2 \pm 4.6 ) \times 10^{-4} \ .
\eeq{epsovereps}

The nonzero result in \leqn{epsovereps} rules out a superweak origin of 
CP violation.  The specific value is too small to be compatible with the
original Weinberg model.  It is an interesting question whether the value
can be compatible with the Kobayashi-Maskawa model or whether it requires new
particles with CP violating couplings.  This topic was discussed at length
at HEP99 \cite{Buchalla},
 and I would like to give my impression of the current situation.

\begin{figure}
\begin{center}
\leavevmode
{\epsfxsize=2.50truein \epsfbox{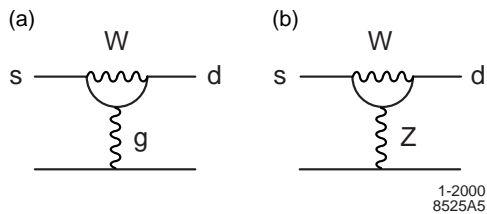}}
\end{center}
\caption{Diagrams contributing to  $\epsilon'/\epsilon$ in 
$K^0$ decays: (a) strong penguin, (b) electroweak penguin.}
\label{fig:penguins} 
\end{figure}

Though the complete formula for $\epsilon'/\epsilon$ in the Standard Model
is very complicated, one can argue about the uncertainties in 
 $\epsilon'/\epsilon$ by using the simplified approximate relation \cite{BB}
\beqa
 \epsilon'/\epsilon &=& 10^{-4} \CR& & 
\hskip -1.5cm \cdot \left[ 16 B_6^{(1/2)}  - 8 
 B_8^{(3/2)} \bigl( {m_t\over 165}\bigr)^{2.5} \right]
 \biggl({110\over m_s}\biggr)^2 \ , 
\eeqa{eoeexp}
where $m_t$ is the $\msbar$ top quark mass evaluated at $m_t$, $m_s$ is the 
$\msbar$ strange quark mass evaluated at 2 GeV, and $B_6{}^{(1/2)}$ and 
$B_8{}^{(3/2)}$ are conventionally defined factors giving the 
matrix elements of penguin operators
arising from the strangeness-changing weak interaction.  The convention for
the $B$ coefficients factors out the dependence on the strange quark mass,
and  one should keep in mind that  it is the combination $B/m_s^2$ which 
corresponds to 
a physical matrix element.  These matrix elements must be determined by a 
nonperturbative technique, for example, lattice QCD simulation.  The first
term in \leqn{eoeexp} is due to the strong-interaction penguin diagram, the
second to the electroweak penguin (as in  Figure~\ref{fig:penguins}).  The
strong cancellation between these two effects for large top quark mass
is the reason that the observed
value of   $\epsilon'/\epsilon$ is much smaller than the original prediction
of Gilman and Wise \cite{FlynnRandall}.  The cancellation also amplifies
the considerable uncertainties in the operator matrix elements.

%%%%%%%%%%%%%%%%%%%%%%%%%%%%%%%%%%%%%%%%%%%%%%%%%%%%%%%%%%%%%%%%%%%%%%%
%%%%%%%%%%%%%%%%%%%%%%%%%%%%%%%%%%%%%%%%%%%%%%%%%%%%%%%%%%%%%%%%%%%%%%%%%%%

New estimates of the parameter $B_8{}^{(3/2)}$  were reported at HEP99:
\beq
B_8^{(3/2)} \cdot  \biggl({110\over m_s}\biggr)^2  = 
          \cases{  0.91\pm 0.16 & \cite{Donoghue} \cr 
                   0.86 \pm 0.07 & \cite{Soni} \cr} \ .
\eeq{Beight}
Both estimates are given in the chiral limit $m_s \to 0$.  (For the 
true value of $m_s$, one should multiply these estimates by 1.3.)
The first of these estimates is based on perturbative QCD analysis of 
spectral-function sum rules, the second is derived from a lattice QCD
calculation using the new technique of domain-wall chiral 
fermions \cite{Kaplan,DWCF,BS}.
From the agreement, it seems that this part of the problem is now fairly 
well understood.  Unfortunately, the situation for  $B_6{}^{(1/2)}$ is much
worse.  This matrix element vanishes in the chiral limit and in the $SU(3)$
limit, making the usual techniques for both lattice gauge theory and QCD
estimates awkward to apply.  For perturbative QCD estimates,   $B_6{}^{(1/2)}$
depends on the scalar and pseudoscalar spectral functions, which are poorly 
known.  The operator which is responsible for the $\Delta I = \half$ rule
enhancement of the $K^0 \to \pi\pi$ matrix element has similar problems, and,
indeed, to this day no lattice gauge theory calculation has been able to 
compute the  $\Delta I = \half$ enhancement accurately.   Thus, the value of
  $B_6{}^{(1/2)}$ is not known, and this uncertainty can easily
allow one to
reconcile the prediction \leqn{eoeexp} with the observed value 
\leqn{epsovereps}.

We have now reached the situation in which we know that CP violation arises
from weak-interaction couplings, but we do not have a sufficiently good 
theoretical understanding of the measured observables to know whether 
CP violation is accounted for by  the Kobayashi-Maskawa model 
or whether new particles with CP violating couplings are required.  
Fortunately, we are entering a new era in which the SLAC, KEK, and Cornell
 $\ee$
B-factories, the HERA-B experiment, and measurements of $B$ decay at
 high-luminosity hadron colliders will provide measurements of new CP violation
observables which can be interpreted with  very small theoretical
uncertainty.  This new era offers us a remarkable opportunity either to 
put the conventional picture of CP violation on a firm footing or to overturn
it and discover signal of new physics.  In order to do this, however, we
must change our view of what the important CP violation observables are and
how we should compare them.

An example of the CP violation observables of the new era is the time-dependent
asymmetry ${\cal A}$ in an exclusive $B$ decay, an observable first
discussed by Carter and Sanda \cite{CSanda}.  For example,
\beq
  \Gamma(\bar B^0 \ \mbox{or} \  B^0 \to J/\psi K^0_S) \sim e^{-\Gamma t}
       \left[ 1 \pm {\cal A} \sin \Delta m_d t\right] \ .
\eeq{Aasym}
In this equation, $\Delta m_d$ is the $B^0_L$--$B^0_S$  mass difference,
and the sign $\pm$ refers to the two possible initial states.
The parameter ${\cal A}$ is manifestly CP violating  and 
can be extracted with essentially no uncertainty from our knowledge of 
hadronic matrix elements.  In the Kobayashi-Maskawa model, ${\cal A} = 
\sin 2\beta$, where $\beta$ is simply related to the phase in the 
Cabibbo-Kobayashi-Maskawa (CKM) mixing matrix.
In models with new CP violating couplings, ${\cal A}$ can obtain additional
large contributions from these sources. 

\begin{figure}
\begin{center}
\leavevmode
{\epsfxsize=3.50truein \epsfbox{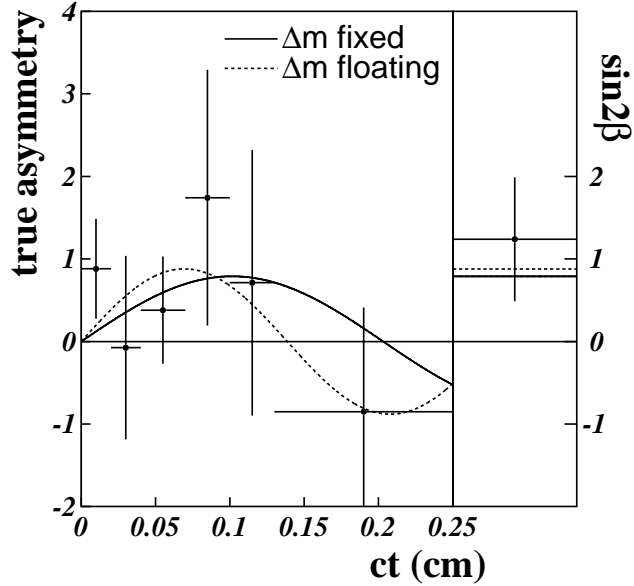}}
\end{center}
\caption{Time-dependent asymmetry in 
$\bar B^0, B^0 \to J/\psi K^0_S$, as measured by the CDF collaboration
[46].}
\label{fig:CDF} 
\end{figure}

At HEP99, we had a first taste of the new era of CP violation with the report
of the first significant measurement the CP asymmetry in $B^0 \to J/\psi K^0$
by the CDF collaboration \cite{CDFpsiK}.  The experiment observed 
the $J/\psi$ in its decay to $\ell^+\ell^-$ and the $K^0_S$ in its decay to 
$\pi^+\pi^-$.  The initial flavor of the $B^0$ was determined either by 
the lepton charge or jet charge on the opposite side of the event, or by 
the charge of a pion accompanying the $B^0$ in the same jet.  The figure of 
merit for such flavor tags, giving the fraction of the event sample that 
corresponds to the effective number of perfectly tagged $B^0$'s, is
\beq
           \epsilon D^2 \ ,
\eeq{figofmerit}
where $\epsilon$ is the efficiency of the tag and $D$ is the dilution, 
the difference between the probability of a correct tag and the probability
of a wrong tag.  (In the next decade, high-energy physicists will mutter
`$ \epsilon D^2$' as often as, in last one, they were heard to mutter
`$\sstw$'.)  For the CDF measurement, the $\epsilon D^2$ for each  of the
three tagging methods
is about 2\%,  so that the sample of 400 events corresponds effectively
to 25 tagged $B^0$ decays.  From this, one finds
\beq
              {\cal A} = 0.79 ^{+0.41}_{-0.44} \ ,
\eeq{CDFpK}
roughly a $2\sigma$ determination that $ {\cal A} > 0$.  I show the data
binned as a function of $t$ in Figure~\ref{fig:CDF} and leave it to you to
judge the quality of the evidence.

%%%%%%%%%%%%%%%%%%%%%%%%%%%%%%%%%%%%%%%%%%%%%%%%%%%%%%%%%%%%%%%%%%%%%%%
%%%%%%%%%%%%%%%%%%%%%%%%%%%%%%%%%%%%%%%%%%%%%%%%%%%%%%%%%%%%%%%%%%%%%%%%%%%

\begin{figure}
\begin{center}
\leavevmode
{\epsfxsize=4.00truein \epsfbox{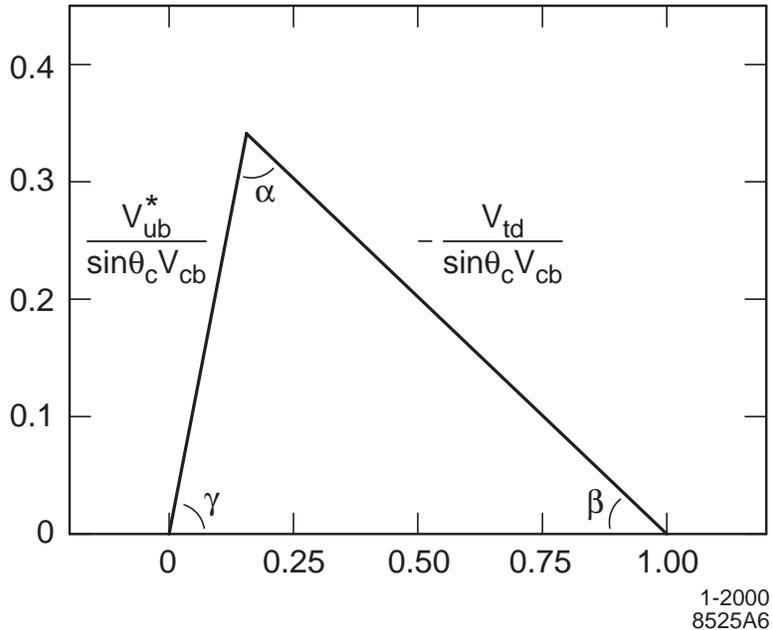}}
\end{center}
\caption{The unitarity triangle.}
\label{fig:Utriangle} 
\end{figure}

Within a year or so, we should have the first accurate measurements of these
new CP observables, and we will need a framework to use in comparing them.
A useful pictorial device is the 
`unitarity triangle' \cite{Bjut,JS}, the
triangle in the complex plane which reflects the unitarity relation of 
CKM matrix elements
\beq
       V_{ub} V_{ud}^* + V_{cb} V_{cd}^* + V_{tb} V_{td}^* = 0 \ ,
\eeq{Bjone}
Using the approximations $V_{ud} \approx V_{tb} \approx 1$, 
$V_{cd} \approx \sin \theta_C$, we find the relation shown in 
Figure~\ref{fig:Utriangle}.  The internal angles of this triangle are referred
to as $\alpha$, $\beta$, $\gamma$, except in the Far East, where the 
notation $\phi_2$, $\phi_1$, $\phi_3$ is used.

%%%%%%%%%%%%%%%%%%%%%%%%%%%%%%%%%%%%%%%%%%%%%%%%%%%%%%%%%%%%%%%%%%%%%%%
%%%%%%%%%%%%%%%%%%%%%%%%%%%%%%%%%%%%%%%%%%%%%%%%%%%%%%%%%%%%%%%%%%%%%%%%%%%

It is often 
said that the goal of the new CP violation measurements is to `check
whether the unitarity triangle closes'.  I would like to substitute for this 
a more precise idea.

Since CP violating phases can be redefined by convention, 
CP violation observables typically involve  phase differences between 
two different amplitudes.  Usually, these are
$B$ or $K$ mixing amplitudes or other loop diagrams on one hand, and 
weak decay amplitudes on the other hand.  I will assume that the phases
of the decay amplitudes come only from the CKM matrix elements.
This is correct unless the decay amplitudes also receive corrections
from the tree-level exchange of light exotic particles such as charged
Higgs bosons.  On the other hand, a loop diagram which contribute to 
mixing can receive corrections from any new particles with masses in the
range up to 1 TeV, and the couplings of these particles can bring new
contributions to its phase.  If we try to determine the unitarity triangle 
from 
a set of processes which all involve the same loop diagram, it is possible
to get a consistently  determined triangle which does not coincide with the
true unitarity triangle of the CKM matrix.  The way to test models of CP
violation, then, is to compare the unitarity triangles determined from 
different classes of CP observables.  This point of view, set out in 
the original work of  Nir and Silverman \cite{NirS}, has been emphasized
more recently by 
Cohen, Kaplan, Lepeintre, and Nelson \cite{CKLN} and
by Grossman, Nir, and Worah \cite{GNW}.

\begin{figure*}
\begin{center}
\leavevmode
{\epsfxsize=6.20truein \epsfbox{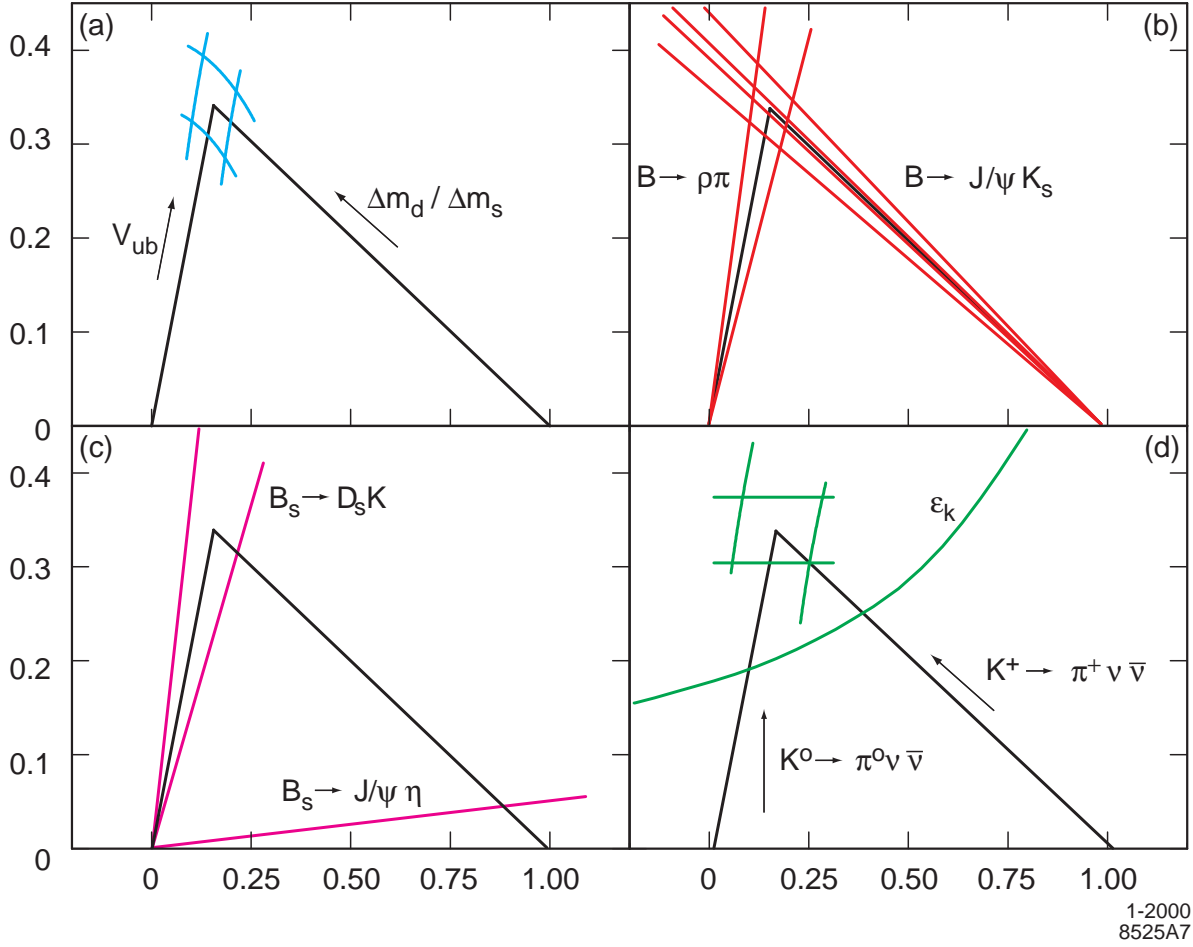}}
\end{center}
\caption{Illustration of four determinations of the unitarity triangle, by 
(a) non-CP observables, (b) $B$ asymmetries, (c) $B_s$ asymmetries, 
(d) $K$ rare decays.  See the text for more details.}
\label{fig:fourtriangles} 
\end{figure*}

I would now like to distinguish four classes of CP violation measurements, 
corresponding to four different physical systems, such that each class
would  determine the unitarity triangle completely
if the Kobayashi-Maskawa model were a complete description of CP violation.
The test of the Kobayashi-Maskawa model will come from the comparison of these
triangles.  The four triangles that I will discuss are shown in 
Figure~\ref{fig:fourtriangles}, with error boxes for the sides or angles
that might be realized within the next decade.

%%%%%%%%%%%%%%%%%%%%%%%%%%%%%%%%%%%%%%%%%%%%%%%%%%%%%%%%%%%%%%%%%%%%%%%
%%%%%%%%%%%%%%%%%%%%%%%%%%%%%%%%%%%%%%%%%%%%%%%%%%%%%%%%%%%%%%%%%%%%%%%%%%%

Figure~\ref{fig:fourtriangles}(a) shows the `non-CP triangle'.  This 
triangle takes advantage of the fact that one can determine the unitarity
triangle by measuring the absolute values of CKM matrix elements and thus
show the existence of the phase through non-CP-violating observables.
The left-hand side of the triangle is determined by the rate of $b\to u$
weak decays; the right-hand side is determined by ratio of $B$--$\bar B$ 
mixing amplitudes for $B_s$ and $B_d$.  The rate of $b\to u$ transitions
depends only on the CKM matrix element $V_{ub}$ and is not affected by 
new physics.  The $B$ mixing amplitudes involve box diagrams that might
have large nonstandard contributions.  However, in many models, including
models with light supersymmetric particles in which squarks with the same
electroweak quantum number are naturally degenerate, these contributions
have the same ratio as the standard contributions \cite{JoAnne}.  Thus, this
`non-CP triangle' is the most likely of the four to agree with the true
unitarity triangle determined from the CKM matrix.

The expected accuracy that I have displayed in this figure---10\% for the 
$V_{ub}$ side and 5\% for the $V_{td}$ side---is surprisingly small, and
I would like to defend these estimates now.  I will begin with $V_{td}$.
This parameter is determined by the relation
\beq
       {\Delta m_d\over \Delta m_s} =
 {m_{Bd} f^2_{Bd} B_{Bd}\over m_{Bs} f^2_{Bs} B_{Bs}} \cdot 
            \left| {V_{td}\over V_{ts}}\right|^2   
 =  \xi^{-1}    \left| {V_{td}\over V_{ts}}\right|^2  \ ,
\eeq{xidef}
where $f_{Bd}$ is the $B_d$ decay constant and $B_{Bd}$ is the matrix element
of a 4-fermion operator in the $B_d$ wavefunction.
The $B_d$ mixing parameter $\Delta m_d$ is now known to  3.5\% 
accuracy \cite{Artuso}.
The $B_s$ mixing parameter can be determined by looking for a fast 
oscillation in tagged $B^0$ decays superimposed on the slow oscillation
from $B_d$ mixing.  There is suggestive evidence that such an oscillation
appears in the $B$ vertex distribution at the $Z^0$, corresponding to an
oscillation frequency $\Delta m_s \sim 16$ ps$^{-1}$ \cite{atLP}; I 
have used this
value in constructing the figure.  Once the oscillation is seen, the frequency
can be determined to a few percent.  The CDF experiment should be able to 
make this measurement early in Run II, even for $\Delta m_s$ so large 
that the triangle
collapses onto the real axis.  Looking back at \leqn{xidef},
the magnitude of $V_{ts}$ is constrained by unitarity to be very close to 
$|V_{cb}|$.  Thus, the main source of uncertainty is in the estimation of 
the $\xi$.  The ratio $\xi^{-1}$ is roughly equal to 0.8 and tends to 1 in the 
chiral limit or in the $SU(3)$ limit $m_d = m_s$.  To achieve 5\% accuracy
in $\xi$, it is only necessary to compute the deviation of $\xi$ from 1 to 
25\% accuracy. 

 Lattice gauge theory should be up to the task.  At HEP99,
the CP-PACS collaboration reported a calculation \cite{CPPACS}
\beq
 \left(      {f^2_{Bd}\over f^2_{Bs}}\right) =  0.69 \cdot 
    (1 \pm 7\% \pm 3\% {}^{+5\%}_{-7\%}) \ ,
\eeq{fratio}
where the three errors come from Monte Carlo statistics, the determination
of $m_s$, and the continuum extrapolation.  It seems to me
that, with further effort,
 a 5\% determination of $|V_{td}/V_{cb}|$ is quite feasible.  A useful 
review of the status of lattice gauge theory determinations of these 
and other heavy-quark matrix elements can be found in \cite{Aoki}.

The situation is less clear for $V_{ub}$, but still there is reason for
optimism \cite{Ligeti}.
  The best current measurement of $V_{ub}$ is based on the 
CLEO measurement of the rate of $B\to \rho \ell \nu$ \cite{CLEOVub},
\beq
    V_{ub} = (3.25\pm 0.14 {}^{+0.21}_{-0.29} \pm 0.55) \times 10^{-3} \ ,
\eeq{VubC}
where the third contribution to the error represents a 20\% spread in the 
relations given by models between the underlying parameters and the observed
rate.  The experimental uncertainties are thus quite adequate, and they will
decrease in the era of the B-factories.  What is needed is a method for 
computing $V_{ub}$ that has less model uncertainty.  Two methods have been
proposed.  The first is an inclusive technique based on the idea that 
in a decay $B\to X\ell\nu$, if $m(X) < m_D$, then the decay must be a 
$b\to u$ transition \cite{BKP,Dai}.  The problem with this method is that
energy from neutral particles cannot be unambiguously associated with a 
displaced vertex, so one must work with vertex masses based on charged
particles and use models to estimate the background from $b\to c$ decays.
The DELPHI collaboration has made a promising first application of this 
technique \cite{DELPHIVub}, obtaining
\beq
       V_{ub} =  ( 4.1 \pm 0.5 \pm 0.6 \pm 0.3 ) \times 10^{-3}\ ,
\eeq{DelphiVub}
where the last error indicates a 8\% model uncertainty.  It is a very 
interesting question how one defines the optimized vertex mass for 
this measurement applicable
to the B-factory environment.  The second method is to measure the spectrum
of $B\to \rho \ell \nu$ decays as a function of $m(\ell\nu)$ and evaluate
it at the `zero-recoil' point where the heavy $B$ quark decays to a $u$
quark at rest.  The value of the form factor at this point can be 
computed by lattice gauge theory simulations \cite{Aoki}.

Figure~\ref{fig:fourtriangles}(b) shows the `B triangle'.  This 
triangle is constructed from the CP asymmetries  in $B^0/\bar{B}{}^0$
decays.  To draw the figure, I have used the asymmetry in $B\to J/\psi K^0_S$
and the asymmetry in $B\to \rho \pi$. (I ignore the discrete ambiguities
in determining the CKM angles from the measured asymmetries.)
Both of these asymmetries involve
the phase in the  $B^0$--$\bar{B}{}^0$ mixing amplitude and are sensitive to 
new physics through this source.
 For  $B\to J/\psi K^0_S$, 
at least four independent experiments (BaBar, BELLE, CDF, HERA-B) should
determine $\sin2\beta$ an accuracy better than $\pm 0.1$ in the near future. 
LHC-B or BTeV should determine this parameter to the level of $\pm 0.01$.
 The constraint from 
 $B\to \rho \pi$ is actually a measurement of $\alpha$ in the CKM picture,
but I have moved the constraint to the lower vertex of the triangle 
for clarity.  The process originally thought to best determine $\alpha$,
$B\to \pi\pi$, is now disfavored due to potential large background 
contributions from 
strong and electromagnetic penguin diagrams.  With  sufficient statistics,
one can fit to the Dalitz plot in $B\to \rho \pi$ to measure and remove
these contributions.  The BaBar collaboration has estimated an accuracy of
$7^\circ$ in $\alpha$ for a  sample of 600 events \cite{BBBook}.  Such a
large sample would require luminosities well above the design level.   It is 
also possible to measure $\gamma$ by the comparison of rates for 
$B^\pm \to K^\pm D$ decays \cite{BKD}.  This determination involves only 
tree-level decay amplitudes and so measures the true CKM unitarity 
triangle rather than  the `B triangle'.

Figure~\ref{fig:fourtriangles}(c) shows the `B$_s$ triangle'.  The 
time-dependent CP asymmetry in $B_s \to D_s^\pm K^\mp$ is connected
to $\sin\gamma$.  LHC-B or BTeV should measure $\gamma$ using this 
reaction to about 10$^\circ$.  The $B_s$ system also allows an interesting
null experiment.  The time-dependent CP violation in 
$B_s \to c \bar c s \bar s$ decays is expected to be very small in the 
Standard Model.  On the other hand, the phase in $B_s \to J/\psi \eta$
and $B_s\to J/\psi \phi$ should be measurable to  a few degrees by 
LHC-B or BTeV.  These reactions will be a very sensitive indicator for
new CP violating physics in the $B_s$--$\bar{B}_s$ mixing amplitude.
This constraint is shown, just for the purpose of illustration, as a 
constraint on the base of the unitarity triangle.

Figure~\ref{fig:fourtriangles}(d) shows the `K triangle'.  This is the 
triangle determined by the rare $K$ decays  $K^+ \to \pi^+ \nu\bar\nu$,
which has an amplitude approximately proportional to $V_{td}$ in the 
Standard Model, and  $K^0_L \to \pi^0 \nu\bar\nu$, which is a CP-violating
process whose amplitude is proportional to Im[$V_{td}$] in the Standard
Model.  These decays proceed through box diagrams which could well have
exotic  contributions from new particles with masses of a few hundred GeV.
The rare $K$  decays are frighteningly difficult to detect.  Experiment
E787 at Brookhaven has recently observed 1 event of the $K^+$ 
decay \cite{Kplus}.  There are preliminary plans for experiments that would
run at Fermilab in the next decay and collect 100 events in each of these
rare modes; I have drawn the triangle assuming that these experiments succeed
and that their statistical errors are dominant.
I have also included in this plot the lower bound of the 
constraint from $\epsilon$, with 
conservative errors \cite{BBBook} reflecting the uncertainty in a lattice 
determination of the overall normalization of  hadronic matrix element.
It should be noted that, while large deviations from the CKM prediction 
are possible in rare $K$ decays, broad classes of models give only  
a relatively small effect \cite{NW}.

\begin{figure}
\begin{center}
\leavevmode
{\epsfxsize=4.00truein \epsfbox{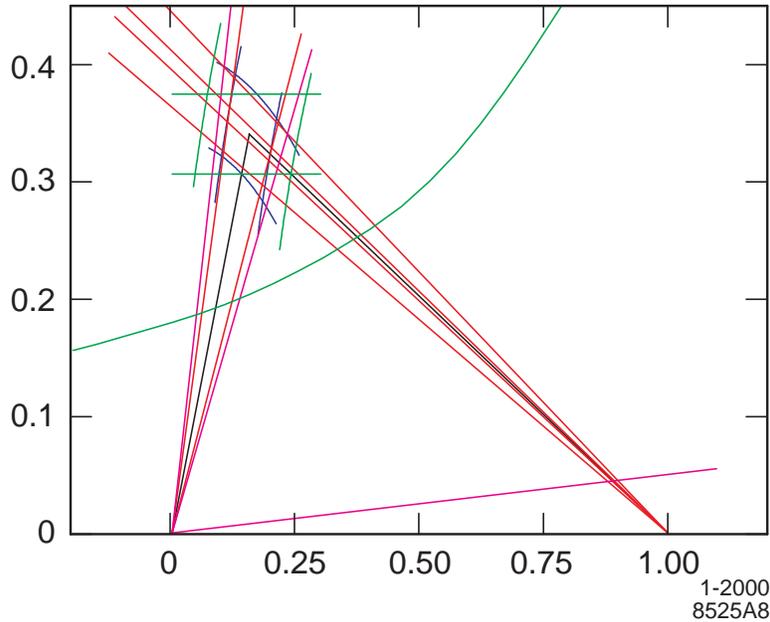}}
\end{center}
\caption{Unitarity triangle determinations from Figure 8 superimposed on a
single set of axes.}
\label{fig:finaltriangle} 
\end{figure}

Figure~\ref{fig:finaltriangle} shows the four unitarity triangles superposed
on one another.  This could well indicate the state of our knowledge of 
CP violation ten years from now.  If the agreement of the various triangles
is as good as what is shown here, it will provide striking evidence that the
Kobayashi-Maskawa model explains the observed CP violation in weak 
interactions.  But keep in mind the possibility that these four triangles
might disagree completely due to loop diagrams involving new heavy 
particles.  We will soon find out which alternative is realized.

%%%%%%%%%%%%%%%%%%%%%%%%%%%%%%%%%%%%%%%%%%%%%%%%%%%%%%%%%%%%%%%%%%%%%%%
%%%%%%%%%%%%%%%%%%%%%%%%%%%%%%%%%%%%%%%%%%%%%%%%%%%%%%%%%%%%%%%%%%%%%%%%%%%

The HEP99 meeting also saw new theoretical developments in the theory of 
$B$ meson CP asymmetries.  In my discussion of $B\to \rho\pi$ above, I
mentioned the difficulties associated with penguin diagrams, which modify
the current-current weak interaction and potentially add a different set of 
phases.  It is an important issue in the theory of $B$ decays to provide 
methods to calculate these penguin contributions or to extract them from data.
I would like to highlight three recent pieces of work along these lines.
In a presentation at HEP99, Fleischer \cite{Fleischer} proposed using 
$SU(3)$ (more 
specifically, U-spin) to relate the decay amplitudes for $B\to \pi^+\pi^-$
and $B_s\to K^+K^-$.  Using these relations, it is possible
to solve for $\beta$  and $\gamma$ without assumptions about the size of 
the penguin effects.  In another recent paper Neubert and Rosner \cite{NR},
following up on ideas of  Fleischer and Mannel \cite{FM}, have shown how
to extract $\gamma$ without assumptions on the size of the penguin 
contributions by fitting all partial rate differences among $B^\pm \to \pi K$
and $B^\pm \to \pi^\pm \pi^0$ decays.  In the most ambitious of these
projects, Beneke, Buchalla, Neubert, and Sachrajda \cite{BBNS}
reported a new factorization
formula  applicable to the decays of a $B$ meson to two pseudoscalars meson
valid in the formal limit $m_b\to \infty$.  In perturbative QCD, the leading
term in this formula is the naive factorization in which one current from the
weak interaction operator creates one final meson.  However, the corrections
to this term are finite and calculable.  A typical result of their method is
the formula
\beqa
   BR(\bar B^0 \to \pi^+\pi^-) &=& 6.5\times 10^{-6}\CR
          &  & \hskip -1.5 cm \cdot 
                    \left| e^{-i\gamma} + 0.09 e^{-i(13^\circ)} \right|^2 \ .
\eeqa{BBNSform}
On the right-hand side, the prefactor should not be taken seriously, but it
can eventually be well determined because it is computed from  
 the same form factor that will be measured in the decay
   $B\to \pi \ell \nu$.  The phase in the 
second term inside the bracket arises from the imaginary part of a 
QCD loop diagram.  It would be 
very interesting to understand the accuracy of the
formulae obtained by this method, since potentially they make many more
processes available for the determination of CKM matrix elements.

Before leaving the subject of CP violation, I would like to remind you that
there are many other possible probes which should be explored.   Even
within the realm of meson asymmetries, there is $D$--$\bar D$ mixing, which
could have a large CP violating component from sources beyond the Standard
Model \cite{NirD}.  Our knowledge of $D$--$\bar D$ mixing will be greatly
improved by the B factory experiments.  Already, CLEO has a new and very
impressive limit, which was presented at HEP99 \cite{CLEODD}.  The neutron
and electron electric dipole moments remain important constraints, especially
on new angles in supersymmetry that couple to light flavors.  CP violation
might also be 
specifically associated with the top quark.  The LHC experiments
should be able to observe
a $10^{-3}$ energy asymmetry between  leptons $\ell^\pm$ produced in top
decays, and this is would be a significant test of CP violation models
\cite{SP,Bern}.  

Finally, it is  important to keep in mind that the 
Kobayashi-Maskawa model of CP violation cannot provide large enough CP 
asymmetries to create the baryon/antibaryon asymmetry in the universe
\cite{HS}.  So, we must eventually find a new source of CP violation.
It is possible that this source is the mass matrix of heavy leptons, or 
some other effect at extremely high energy.  But it is also possible
that the new mechanism of CP violation will appear in the experimental 
program that we are now beginning to carry out.

\section{QCD}

We turn next to Quantum Chromodynamics (QCD). 
 In contrast to the previous two topics, the fundamental
questions about QCD  have been
answered already some time ago.  The experimental confirmation of QCD as the
theory of the strong interactions is now very strong.  QCD is now known to 
account for a wide variety of 
processes with large momentum transfer, in $\ee$ annihilation, $ep$ collisions,
and $p\bar p$ collisions, to about the 10\% level of accuracy, and with a 
common value of the coupling constant $\alpha_s$.  At the same time, numerical
lattice studies confirm that QCD explains the spectrum of light hadrons to
 about the same level of numerical precision.  Wittig has reviewed this latter,
less familiar, evidence for QCD at HEP99 \cite{Wittig}.
So,  what are the important scientific issues for QCD today?  I would like to
highlight four of these, and give some examples of new work presented at HEP99.

%%%%%%%%%%%%%%%%%%%%%%%%%%%%%%%%%%%%%%%%%%%%%%%%%%%%%%%%%%%%%%%%%%%%%%%
\begin{figure*}
\begin{center}
\leavevmode
{\epsfxsize=5.00truein \epsfbox{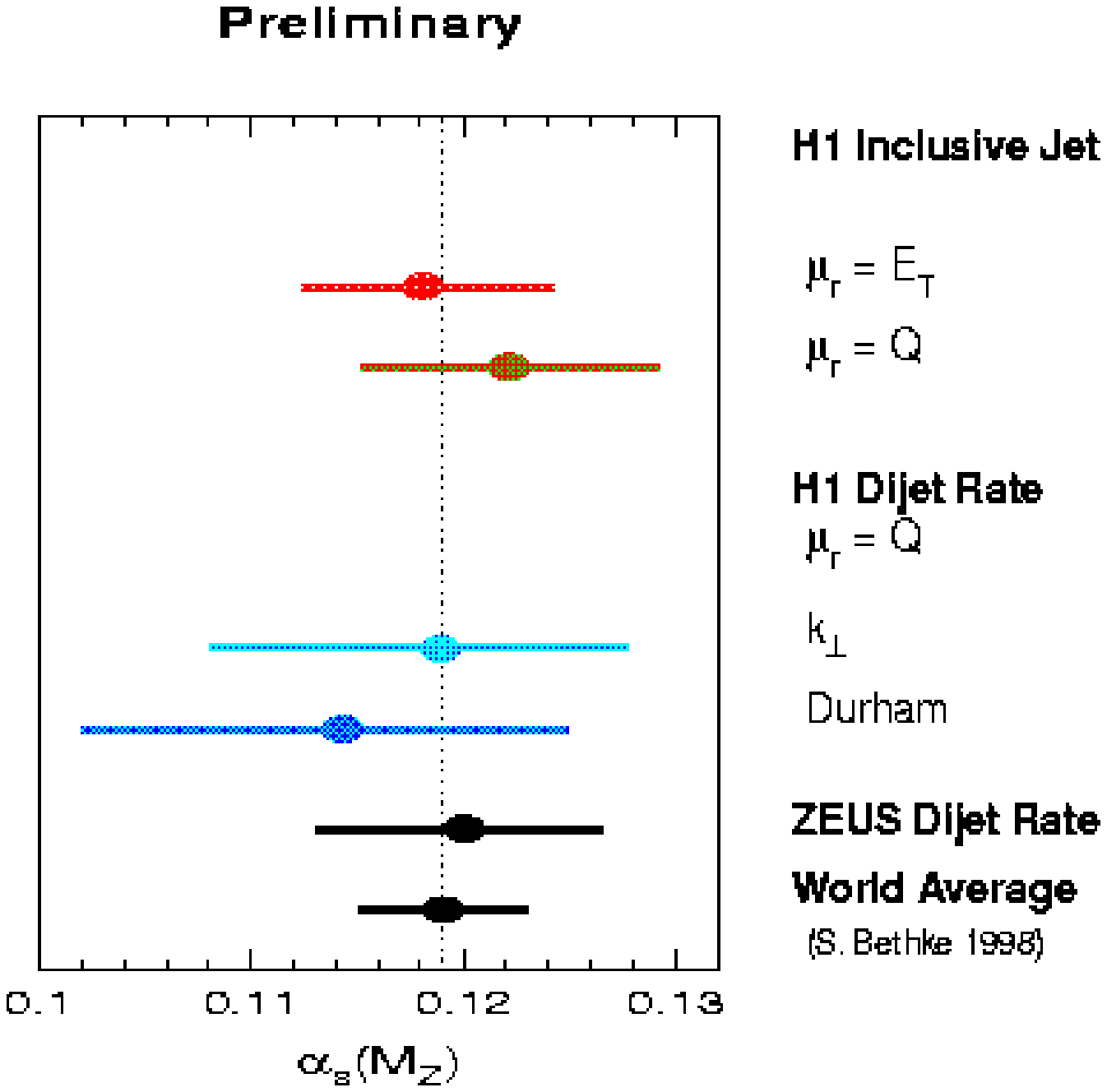}}\hspace*{-1.5in}
{\epsfxsize=2.00truein \epsfbox{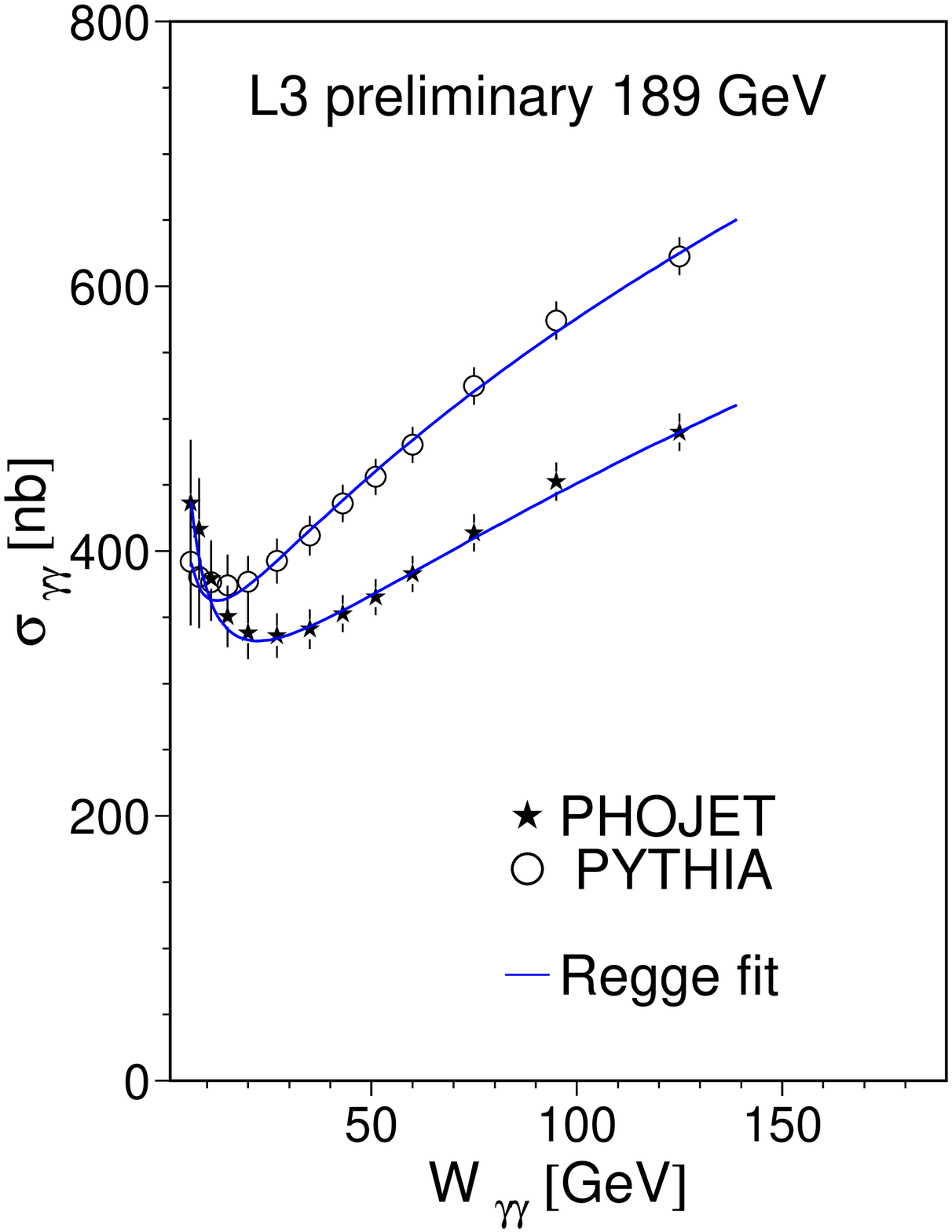}}
{\epsfxsize=2.20truein \epsfbox{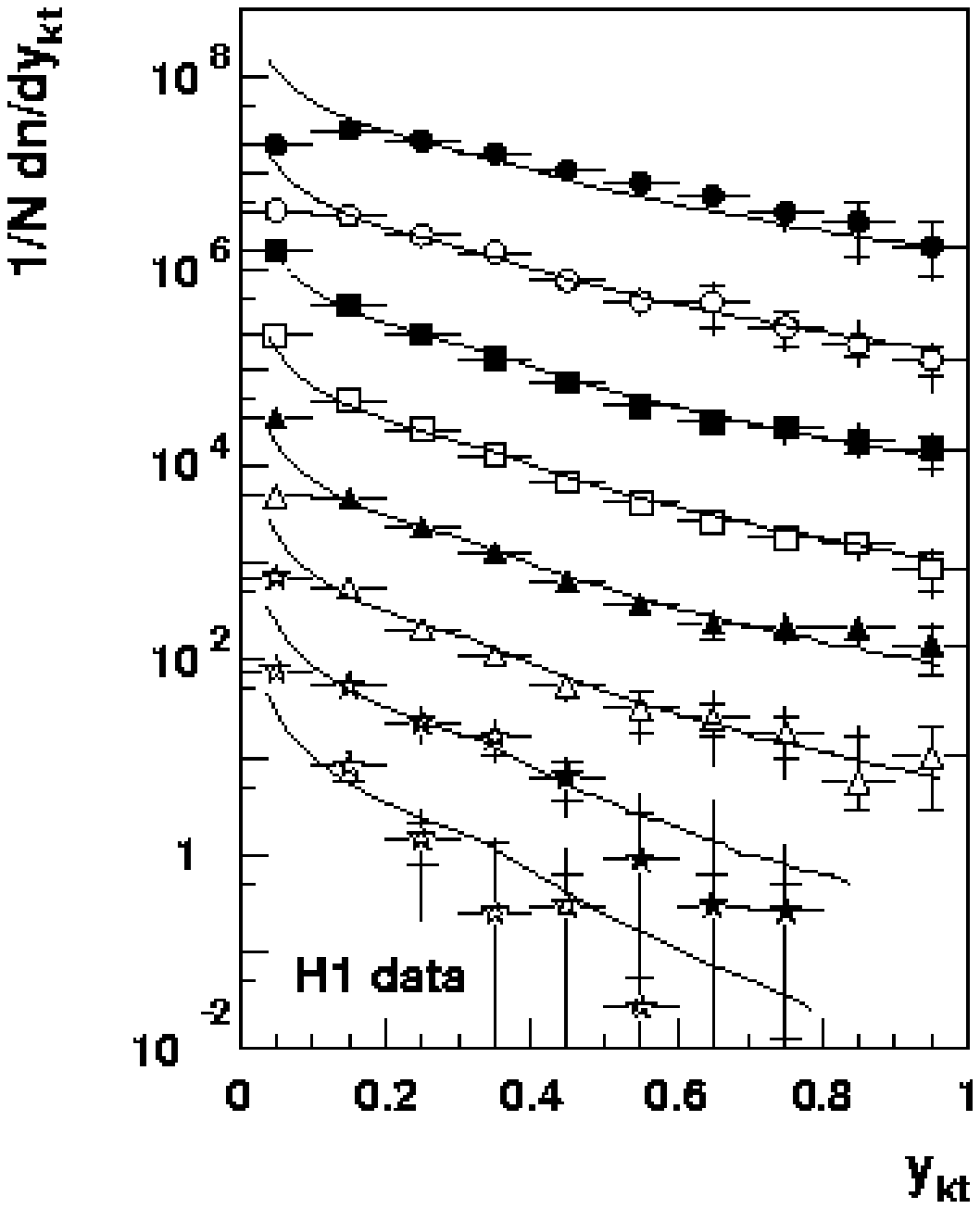}}\hspace*{0.6in}
{\epsfxsize=2.50truein \epsfbox{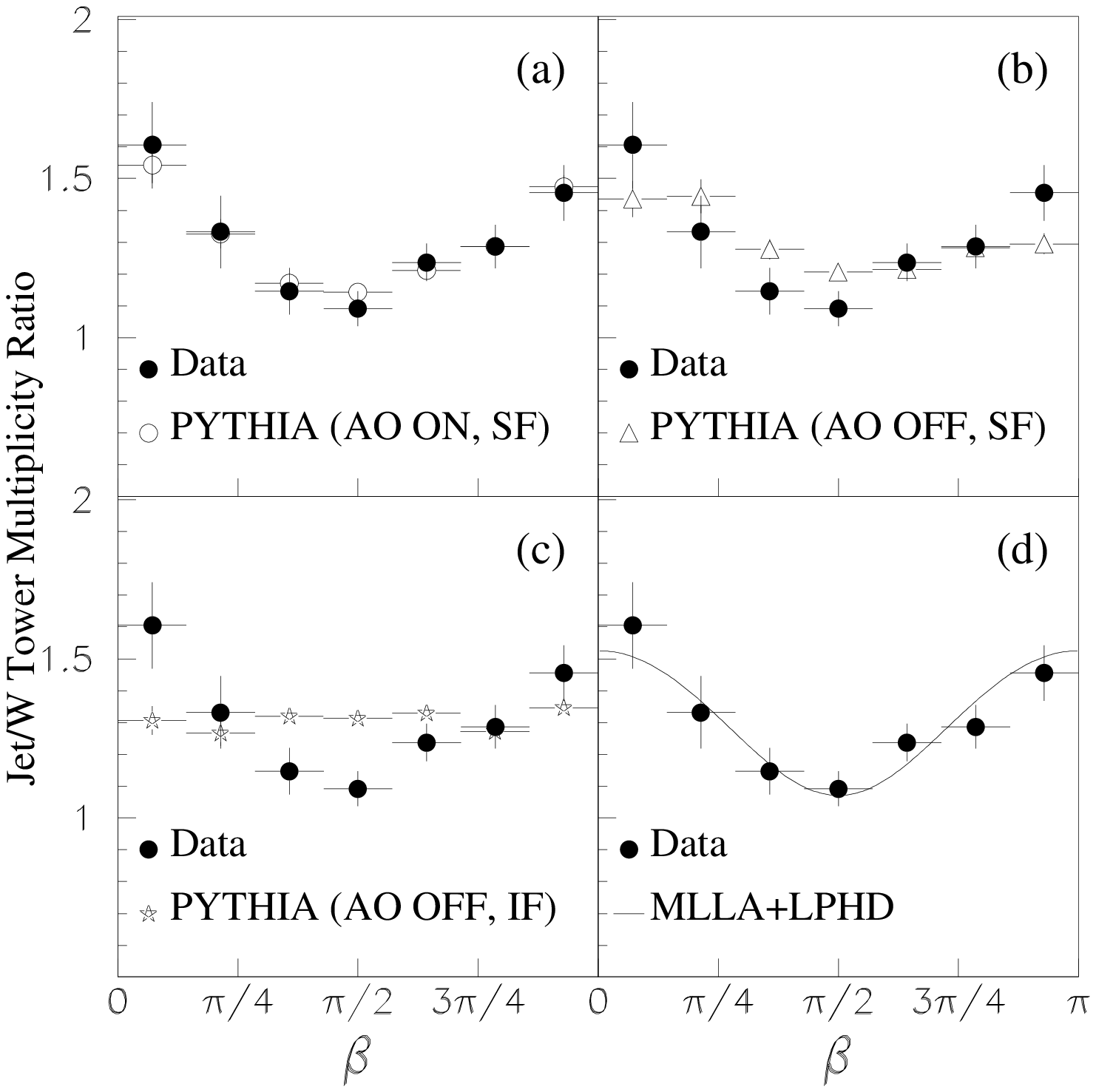}}
\end{center}
\caption{Four figures which show recent experimental advances in QCD.
 See the text for more details.}
\label{fig:montage2}
\end{figure*}
%%%%%%%%%%%%%%%%%%%%%%%%%%%%%%%%%%%%%%%%%%%%%%%%%%%%%%%%%%%%%%%%%%%%%%%%%%%

The first issue is the precision determination of $\alpha_s$.  At the moment, 
the $\msbar$ coupling
$\alpha_s(\mz)$ is known to about 3\% accuracy \cite{alphas}.  It is important
to reduce this error below 1\%.  This level of accuracy is needed as an 
input to the precision experiments of the next decade, for example, the 
study of the top quark at threshold.  It is also already needed to assess the
validity of grand unification.  I have already noted that grand unification
with the renormalization group equations of supersymmetry successfully
relates the values of the Standard Model couplings given in \leqn{SMcouplings}.
In particular, the prediction for $\alpha_s$ agrees with experiment at the 
10\% level, but it is subject at this level to uncertainties from threshold
corrections at the scale of grand unification.  With a more accurate 
$\alpha_s$,  we could evaluate the needed threshold contribution and begin
to test explicit models of grand unification.  The primary barrier to a
more accurate determination of $\alpha_s$ come not from experiment (though
it would be good to have more precise data on multi-jet rates at energies
well above the $Z^0$) but rather from theory.  However difficult it may be,
we need the order $\alpha_s^2$ corrections to the most important processes
which determine $\alpha_s$, in particular, the rate for $\ee \to 3$ jets. 

The second issue is the determination of essential strong interaction 
parameters needed for high energy experiments.  Here I mean especially
the parton distributions in the proton.  Though the quark distributions are
well determined, the gluon distribution is not well constrained at moderate
values of $x$.  This is the freedom that was used to correct the discrepancy
between the CDF measurement of the jet rate at large $E_T$ and QCD predictions
\cite{fixjet}.  Gluons at moderate $x$ and low $Q$ evolve to the gluons 
at low $x$ and high $Q$ which are the dominant source of new particle 
production at the LHC.  The study of high-energy $\ee$ collisions requires
another set of input data, the total cross section for 
the process $\gamma\gamma \to$ hadrons.  It is interesting in its own right
to understand what part of this cross section comes from pointlike processes
and what part from soft processes involving the hadronic constituents of the
photon.  The eventual theory should explain, as be constrained by, the data
both for $\sigma(\gamma\gamma)$ and $\sigma(\gamma p)$.

The third issue is the study of the detailed structure of jets as predicted
by QCD.  QCD predicts that the hardest components of 
jets are built up by successive processes in which 
gluons or quarks split off from the hardest parton.  This gives jets a 
fractal structure.  On top of this backbone, hadrons are produced in a 
way that reflects the color pairings of the hard partons.  If groups of 
partons are separated in phase space and are separately color neutral, we
should find a phase space or rapidity gap in particle production.  All of these
features are just beginning to be understood from the data.

The fourth issue is one for theorists, the development of new techniques to
compute higher-order and multiparton QCD amplitudes.  This issue provides
essential theoretical support to the first and third topics just listed.
At HEP99, Uwer \cite{Uwer}, Draggiotis \cite{Drag}, and Harlander \cite{Harl}
set out new ideas for calculational programs whose results should be very
interesting.

As an illustration of recent progress in these areas, I present in
 Figure~\ref{fig:montage2} four of the new QCD results presented at HEP99.
The upper left \cite{alphafromH1}
shows that event shape measurements from HERA
are now contributing to the precision $\alpha_s$ determination.  
The upper right \cite{costantini} shows the new determination
by L3
of the $\gamma \gamma$ total cross section at LEP. (A similar determination
was presented for the  $\gamma p$ total cross section 
at HERA \cite{wodarczyk}.)
The large uncertainty,
reflected in the difference between the two fits, comes from the fact that 
about 40\% of the total cross section is unobserved, and that theoretical
models differ on the size of the  contribution from these very soft events.
This is a problem that must be addressed.  The last two figures show
new studies of QCD event shapes.  
The lower left \cite{berger} shows the narrowing of jets
with increasing $Q^2$ in  $ep$ collisions in the H1 event sample.
The lower right, from the D0 experiment \cite{varelas},
 shows that the particle production
in $W +$ jet events reflects the color flow expected for a color singlet $W$
recoiling against a colored parton.

\section{Supersymmetry}

From the areas of current experimental interest, we now turn to the future.
What should be the main topic of discussion at HEP09?  What new era in 
experimental high energy physics will be opening up at that time?

In this section, I would like to take very seriously the second 
moral I drew in 
Section 2 from the precision electroweak data.  The precision study of the
$Z^0$ points us toward a world in which the interactions responsible for 
electroweak symmetry breaking are weakly coupled and the Higgs boson is
an elementary scalar particle.  I have already explained that some 
specific aspects
of the data are compatible with the idea of supersymmetry at the weak 
interaction scale.  But there is a much stronger argument for the presence
of supersymmetry in the fundamental description of Nature.  While it is 
possible in principle that there is no explanation for the negative
value of the Higgs (mass)$^2$ and the instability to symmetry breaking
in the Higgs potential, I insist that we must find a way to {\em explain}
this instability on the basis of physics.  For this, we must have a 
theoretical framework for the Higgs field in which its potential energy 
function is calculable.  A part of this requirement is that some symmetry
must forbid the addition of a Higgs mass term by hand.  If  the 
Higgs boson is treated as an elementary field, the only known symmetry
with this power is supersymmetry.  Thus, to the extent that the precision
electroweak data excludes models in which the Higgs is composite or 
strongly coupled, we should expect to see not a light Higgs boson but also
the new particles predicted by supersymmetry.

The idea that the data drives us to a weak-coupling picture of the Higgs 
boson was controversial at HEP99.  Among the people arguing vocally on the
other side were Holger Nielsen and Gerard 't Hooft.  So if you do not wish
to accept this argument, you are in good company (but still wrong).
In any event, for the rest of this lecture I will take this conclusion very
seriously and use it to map out future questions for high-energy 
experimentation.

Among the many theoretical problems connected with supersymmetry, I would 
like to focus on the spectrum of supersymmetric particles.  Supersymmetry
predicts that every particle of the Standard Model has a partner with the 
opposite statistics.  That is, the chiral fermions of the Standard Model
have scalar partners, and the gauge bosons have spin-$\half$ (`gaugino')
partners.  What are the masses of these particles?  

The interest of this 
question goes beyond the issue of where or when these particles will be 
found. To produce a reasonable phenomenology, supersymmetry must itself be
spontaneously broken.  The supersymmetrized Standard Model cannot directly 
break supersymmetry, because this hypothesis leads to unwanted very light
superpartners \cite{DG}.  In most models, the description of 
the superparticle masses involve two ingredients, a sector in which 
supersymmetry is broken and a `mediator' which connects the symmetry-breaking
to the Standard Model fields.  The identity of the mediator is typically
connected to the very short distance physics of the model, the connection
of the Standard Models fields to grand unification or to gravity. And,
the nature of the mediator is reflected in the 
detailed pattern of the masses of superparticles.  

If supersymmetry explains the
phenomenon of electroweak symmetry breaking, the superparticle masses must
have masses of a few hundred GeV; that is, they should be accessible to the 
experiments of the next decade and possibly even to LEP and the Tevatron.
By measuring this mass spectrum, we should,
ten years from now, have a wealth of new data which speaks directly
to the physics at this fundamental level.  It is an exciting prospect.
To help you think about it more clearly, I would like to offer a first lesson
in superspectroscopy.  I will contrast three paradigmatic theories of the 
mediation of supersymmetry breaking.  Though supersymmetry phenomenology 
has been studied for almost twenty years and went through a period of 
complacency in the early 1990's, two of these paradigms were discovered 
only recently.  The 
second was invented in  1995, and the third only in the past year.  Presumably,
there are more theoretical insights into this subject that are
 waiting to be uncovered.

%%%%%%%%%%%%%%%%%%%%%%%%%%%%%%%%%%%%%%%%%%%%%%%%%%%%%%%%%%%%%%%%%%%%%%%%%
\begin{table*}
\begin{center}
\caption{Basic formulae for superparticle masses in three paradigms for the
spectrum.  In this table, $i=1,2,3$ refers to a factor of the gauge group
$U(1)\times SU(2)\times SU(3)$, $\alpha_i$ is the gauge coupling evaluated
at $\mz$, $\alpha_U$ is the unification coupling, 
$C_i$ is the squared gauge charge, and $f$ refers to a chiral 
quark or lepton flavor.
The renormalization group functions of the 
supersymmetric Standard Model are notated:  $\beta_i = - b_i g_i^3/(4\pi)^2$,
$\gamma_f = - \eta_{fi} g_i^2/(4\pi)^2$.  In the gauge mediation line, the
factor 2 in the scalar mass formula is actually model-dependent.}
\bigskip
\begin{tabular}{lcc} 
\hline
      &    Gaugino $m^2$      &       Scalar $m^2$ \\
\hline\hline
Gravity    & $\displaystyle{ {\alpha_i^2\over \alpha_2^2} m_2^2}$  & 
  $ \displaystyle{  m_0^2 + \sum_i 2 C_i 
  {\alpha_i^2-\alpha_U^2\over b_i\alpha_2^2} m_2^2}$ \\
Gauge      &  $ \displaystyle{{\alpha_i^2\over \alpha_2^2} m_2^2}$  & 
                      $\displaystyle{\sum_i 2 C_i 
  {\alpha_i^2\over \alpha_2^2} m_2^2}$ \\
Anomaly   &    $\displaystyle{{b_i^2\over b_2^2} {\alpha_i^2\over \alpha_2^2}
                      m_2^2}$  & 
    $\displaystyle{\sum_i 2 \eta_{fi} b_i 
                     {\alpha_i^2\over \alpha_2^2} m_2^2}$ \\
\hline
\end{tabular}
\end{center}
\label{tab:Super}
\end{table*}
%%%%%%%%%%%%%%%%%%%%%%%%%%%%%%%%%%%%%%%%%%%%%%%%%%%%%%%%%%%%%%%%%%%%%%%%%%%

The three paradigms for the superspectrum that I would like to discuss are
those of `gravity mediation', `gauge mediation', and `anomaly mediation'.
In gravity mediation, as introduced in \cite{grav1,grav2,grav3}, the 
mediator is supergravity itself.  One imagines a sector that spontaneously 
breaks supersymmetry.  Let $m_\Pl$ be the Planck scale. Then 
  the gaugino masses arise from direct 
order $1/m_\Pl$ couplings of this sector to the Standard Model
Yang-Mills Lagrangian, and scalar masses arise from direct order $1/m_\Pl^2$ 
couplings of this sector into the Standard Model potential.  If the 
superparticle masses are of the order of the weak scale, the mass of the 
gravitino is of the same order.  The spectrum
acquires additional structure from the renormalization group evolution of the
mass parameters from the Planck scale to the weak scale.  

In gauge mediation, introduced in \cite{DNNS}, the mediator is the 
supersymmetric version of the Standard Model gauge interactions.  One 
imagines that supersymmetry is broken by a new sector which includes 
heavy particles
with Standard Model gauge couplings.   Then gaugino masses arise from 
one-loop diagrams involving these heavy particles, and scalar masses 
arise from 
two-loop diagrams  in which the loop of heavy particles appears in a 
scalar self-energy diagram.  Because what is computed for scalars is 
the (mass)$^2$, both gauginos and scalars acquire masses of order $\alpha$
with respect to the heavy scale.  The gravitino has a mass of order eV and
is typically the lightest superparticle.  Other superparticles may be 
observed to decay to the gravitino if the rate of this decay is not 
suppressed by a very large value of the heavy mass.

Anomaly mediation, introduced in \cite{LGMR,RSanom}  represents the opposite
extreme pole.  To realize this possibility, one considers models in which 
the supergravity couplings needed to work gravity mediation vanish.
Then the partners of Standard Model particles acquire no mass at tree 
level.  In this case, it turns out that the first mass contribution is
universal in character and is connected to the breaking of scale invariance
by the running of the Standard Model coupling constants.  The gravitino
mass does arise at the tree level, and so in this scheme the gravitino mass
is about 100 TeV if the Standard Model superparticles acquire weak scale
masses.

In Table~\ref{tab:Super}, I have collected the basic formulae for the 
gaugino and scalar masses in these three paradigms.  For clarity, I have
eliminated the underlying parameters in terms of the mass $m_2$ of the 
$W$ boson superpartner.  In the case of gravity mediation, there is 
another independent parameter $m_0$.   For gauge mediation and anomaly 
mediation, the superparticle masses naturally depend only the Standard 
Model quantum numbers, a feature that suppresses new flavor-changing 
neutral current effects from supersymmetry loop diagrams.  This property
also holds in gravity mediation in the `no-scale' limit $m_0\to 0$.
Away from this limit, there is no clear reason why $m_0$ should not
depend on flavor except that this could lead to unwanted neutral current
effects.

  Anomaly mediation makes two specific predictions that deserve
comment.  First, the lightest gaugino are expected to be an almost 
degenerate $SU(2)$ triplet $\s w^\pm$, $\s w^0$, with the charged states only
a few hundred MeV above the neutral one \cite{WT}.  This gives rise to a 
distinctive phenomenology, discussed in \cite{Feng,GWells,GM}.  Second, the 
scalars partners of leptons are computed to have negative (mass)$^2$, a 
disaster.  Some cures for this problem are given in \cite{Pomarol,Shadmi}.

\begin{figure*}
\begin{center}
\leavevmode
{\epsfxsize=5.00truein \epsfbox{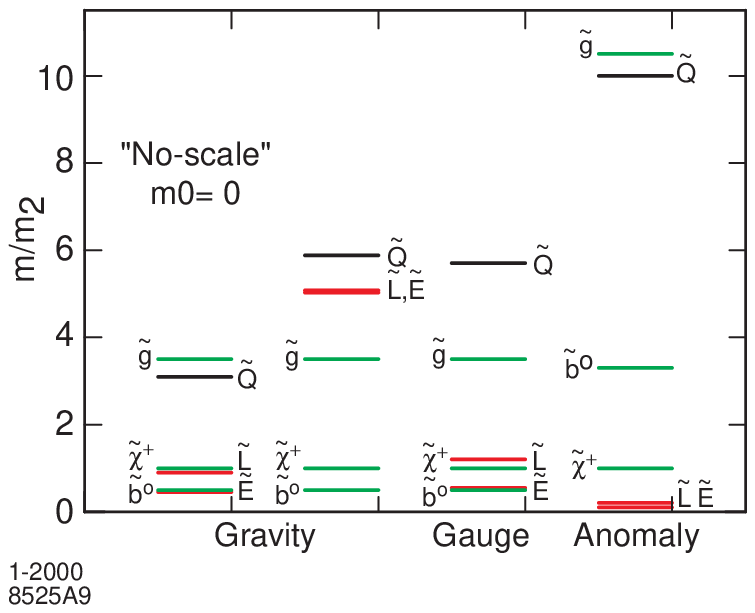}}
\end{center}
\caption{Spectra of superparticles in two cases of gravity mediation, 
gauge mediation, and anomaly mediation, according to the formulae of 
Table 1.}
\label{fig:Super}
\end{figure*}

  In Figure~\ref{fig:Super}, I show a comparison of the
 spectra for the three paradigms; for gravity mediation, I give both 
the `no-scale' case  $m_0 = 0$ and the case $m_0/m_2 = 5$ and universal.

%%%%%%%%%%%%%%%%%%%%%%%%%%%%%%%%%%%%%%%%%%%%%%%%%%%%%%%%%%%%%%%%%%%%%%%
%%%%%%%%%%%%%%%%%%%%%%%%%%%%%%%%%%%%%%%%%%%%%%%%%%%%%%%%%%%%%%%%%%%%%%%%%%%

I emphasize that the
formulae in Table~\ref{tab:Super}
represent only the first lesson in superspectroscopy.  They omit
possible mass mixings and effects of large top, bottom, and $\tau$
Yukawa couplings, and they omit higher-order corrections \cite{bagger}.
Nevertheless, these formulae and Figure~\ref{fig:Super} 
already give a feeling for the complexity of the spectrum that might
be found when superparticles appear in experiments.

If supersymmetry is the explanation of electroweak symmetry breaking, it is
likely that the LHC will be able to sample the whole superparticle mass 
spectrum, including the heaviest states.  In addition, as 
members of the ATLAS collaboration have recently demonstrated, the 
LHC experiments have  the ability 
to make precision measurements of superparticle masses
in a number of different scenarios \cite{Hinch}.  Nevertheless, 
the richness of the phenomena calls for the exploration of 
these particles also in $\ee$ annihilation.  It is worth remembering that
an $\ee$ linear collider of the next generation will provide not only a 
relatively clean environment with kinematic constraints that aid in particle
mass measurements, but also the availability of beam polarization, which is
very useful in resolving questions of particle mixing \cite{Peskin}.
  The production cross
sections for superparticles are electroweak and can be computed precisely,
allowing an unambiguous determination of the quantum numbers of each new
particle.

The superspectrum is complex, but the LHC and linear collider are powerful
instruments.  With these two facilities, with their complementary strengths,
we could fully explore the supersymmetry spectrum of particles and mine
the information it contains for information about a truly fundamental 
level of physics.  This is already cause for optimism about the future of
experimental particle physics.  But, there is more.

\section{New space dimensions}

Many people say that the key problem of quantum physics has nothing to do 
with what we do at accelerators.  Rather, they say, it is the problem of the
compatibility of  quantum mechanics with general relativity.  To solve this 
problem, one must do two things, first, remove the divergences from the 
quantum theory of gravity, and, second, unify gravity with the microscopic
particle interactions.

Most people who recite this litany do not realize that we have at least
one possible solution already in hand.  It is string theory.  String theory
has not been proved to be the correct theory of Nature, but it does 
demonstrably solve these two problems.  It is the only known approach to these
problems which has no glaring weaknesses.  Therefore, we must take it 
very seriously. 

There has been tremendous progress in string theory since the 1995 discovery
of  string dualities by Hull and Townsend \cite{HT} and Witten \cite{Witt}.
Just in the past year, two new and very profound ideas have been validated:
The first is the idea of `t Hooft \cite{tHholo} and Susskind
\cite{Suss} that quantum gravity is `holographic', in the sense that its
physical degrees of freedom are those of a manifold with one lower dimension
that the dimension of space time.  The second is an explicit realization 
of this relation, due to Maldacena \cite{Malda}, a duality linking
  supersymmetric
Yang-Mills theory in 4 dimensions with supergravity in 5-dimensional 
anti-de Sitter space. I do not have space here to do justice
to these ideas, but they are described clearly in Bachas' lecture at 
HEP99 \cite{Bachas}.

I would like to concentrate instead on another consequence of the new 
understanding of string dualities. These developments have led to new
classes of models in which quantum gravity and string physics is much more
accessible to experiment and may even appear directly in the realm of the 
LHC and the linear collider.

String theory requires that we live in a world which has 11 dimensions.  Until
recently, it was thought that this could only be compatible with our 
observations if seven of these dimensions were compact and very small, of the
order of the Planck scale.  It is interesting, though, to think about new
space dimensions that are not so small.  In 4 dimensions, the gravitational
force falls off as
\beq
       F \sim {m_1 m_2 \over r^2} \ ;
\eeq{fourforce}
in  $(4+n)$ dimensions, it falls off as
\beq    
       F \sim {m_1 m_2 \over r^{2+n}} \ .
\eeq{nforce}
Consider the $n$ extra dimensions to be periodic with period $2\pi R$.  Then
two masses separated by a distance much larger than $R$ would feel a 
gravitational force of the form \leqn{fourforce}, while two masses separated
by a distance much less than $R$ would feel a gravitational force of the 
form \leqn{nforce}.  The short-distance force is the more fundamental.
We can define the fundamental quantum gravity scale $M$ by writing the 
dimensionful constant of proportionality in \leqn{nforce} as a numerical 
constant times $M^{-(n+2)}$.  The constant of proportionality 
in \leqn{fourforce} is just Newton's constant.  If we insist that the forces
match at the distance scale $R$, we obtain the relation
\beq
         (4\pi G_N)^{-1} =  R^n M^{n+2} \ .
\eeq{RMGrel}
This equation has a surprising implication.  If we fix $G_N$ to its observed
value and imagine larger values of $R$, then the true fundamental quantum
gravity scale becomes smaller.  

\begin{figure}
\begin{center}
\leavevmode
{\epsfxsize=2.00truein \epsfbox{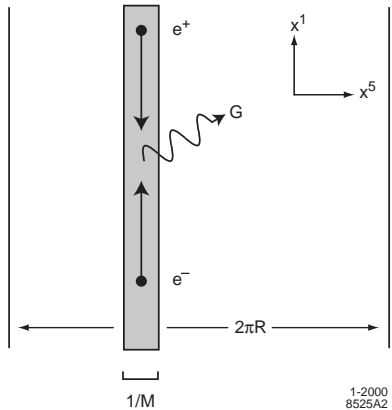}}
\end{center}
\caption{Picture of the universe viewed in the small, with quarks, leptons,
and gauge fields bound to a D-brane localized in an extra compact dimension.}
\label{fig:brane}
\end{figure}

How large could $R$ be?  In principle, $R$ could vary continuously.  But there
are four natural choices represented in the literature as explicit classes of
models.  Using a standard American nomenclature, these sizes are:
\begin{enumerate}
\item {\bf micro-}:  In this case, all three quantities in \leqn{RMGrel} are
of the order of $M_\Pl$.  This is the original proposal of Scherk and Schwarz
for the size of extra dimensions in string theory \cite{SS}.
\item {\bf mini-}:  In this case, $M$ is taken to be of the order of the 
grand unification scale, $2\times 10^{16}$ GeV.  
In fact, Ho\v rava and Witten \cite{HW}
have argued that there is a solution in which all fundamental scales in 
Nature are of order the grand unification scale, with the scale of the 
11th dimension only a small amount larger. This theory includes the
unification of Standard Model couplings provided by supersymmetry, and 
a unification with gravity as well.  
\item {\bf midi-}:  In this case, $R$ is taken to be at the TeV scale.  From
\leqn{RMGrel}, for $n=6$ for example, the fundamental gravity scale $M$ would
be 8000 TeV.  This choice was first advocated by Antoniadis \cite{Anto},
who showed how it could lead to superparticle masses of the order of the 
weak interaction scale.  It is possible to arrange a unification of Standard
Model couplings at the scale $M$ \cite{Gherg}.  Recently, Randall and 
Sundrum \cite{warp} 
have presented a 5-dimensional model with curvature in the extra 
dimension which gives a novel way to relate the Planck scale and the TeV
scale.  The phenomenology of this model is quite similar to that of models
with flat extra dimensions of TeV-scale size \cite{warp2}.
\item {\bf maxi-}:  In this case, $M$ is  taken to be at the TeV scale.
Then $R$ would be at some scale from millimeters ($n=2$) to fermi ($n=7$).
These large distances, which are huge on the scale of high-energy physics,
would seems to violate common sense. But Arkani-Hamed, Dimopoulos, and Dvali
\cite{ADD} have argued that this aggressive choice is  not excluded.  In this
case, the quantum gravity scale $M$, the shortest possible distance in Nature,
might already be accessible to accelerator experiments.
\end{enumerate}
The maxi- case requires one extra condition.  From tests of Bhabha scattering
and fermion pair production at LEP and quark-antiquark scattering at the 
Tevatron, we know that the strong, weak, and electromagnetic interactions
follow the force laws predicted for four dimensions up to momentum transfers
of about 1 TeV.  This means that the quarks, leptons, and gauge bosons must be
confined to a 4-dimensional submanifold of thickness less than 1/TeV inside
the new large dimensions.  This is known to be possible in string theory.
Indeed, string theory contains a classical solution called a `D-brane', which
can have fermions, bosons, and gauge fields bound to its 
surface \cite{Polch,Polch2}.  In the
maxi- case, the Standard Model particles would live on a D-brane, whose 
thickness would be of order $1/M$, while gravity and perhaps other light
fields could propagate in the full space, out to distances of order $R$.
A picture of this construction is given in Figure~\ref{fig:brane}.

%%%%%%%%%%%%%%%%%%%%%%%%%%%%%%%%%%%%%%%%%%%%%%%%%%%%%%%%%%%%%%%%%%%%%%%
%%%%%%%%%%%%%%%%%%%%%%%%%%%%%%%%%%%%%%%%%%%%%%%%%%%%%%%%%%%%%%%%%%%%%%%%%%%

The micro- case above is famously difficult to test and has given rise to 
unfortunate statements that string theory is not physics.  But I would like
to argue now that the other three cases are amenable to experimental 
test, and that in fact they can be tested at the LHC and the linear 
collider.  Indeed, ten years from now, we could be arguing from experimental 
data about the true number of space dimensions in Nature.

\begin{figure}
\begin{center}
\leavevmode
{\epsfxsize=4.90truein \epsfbox{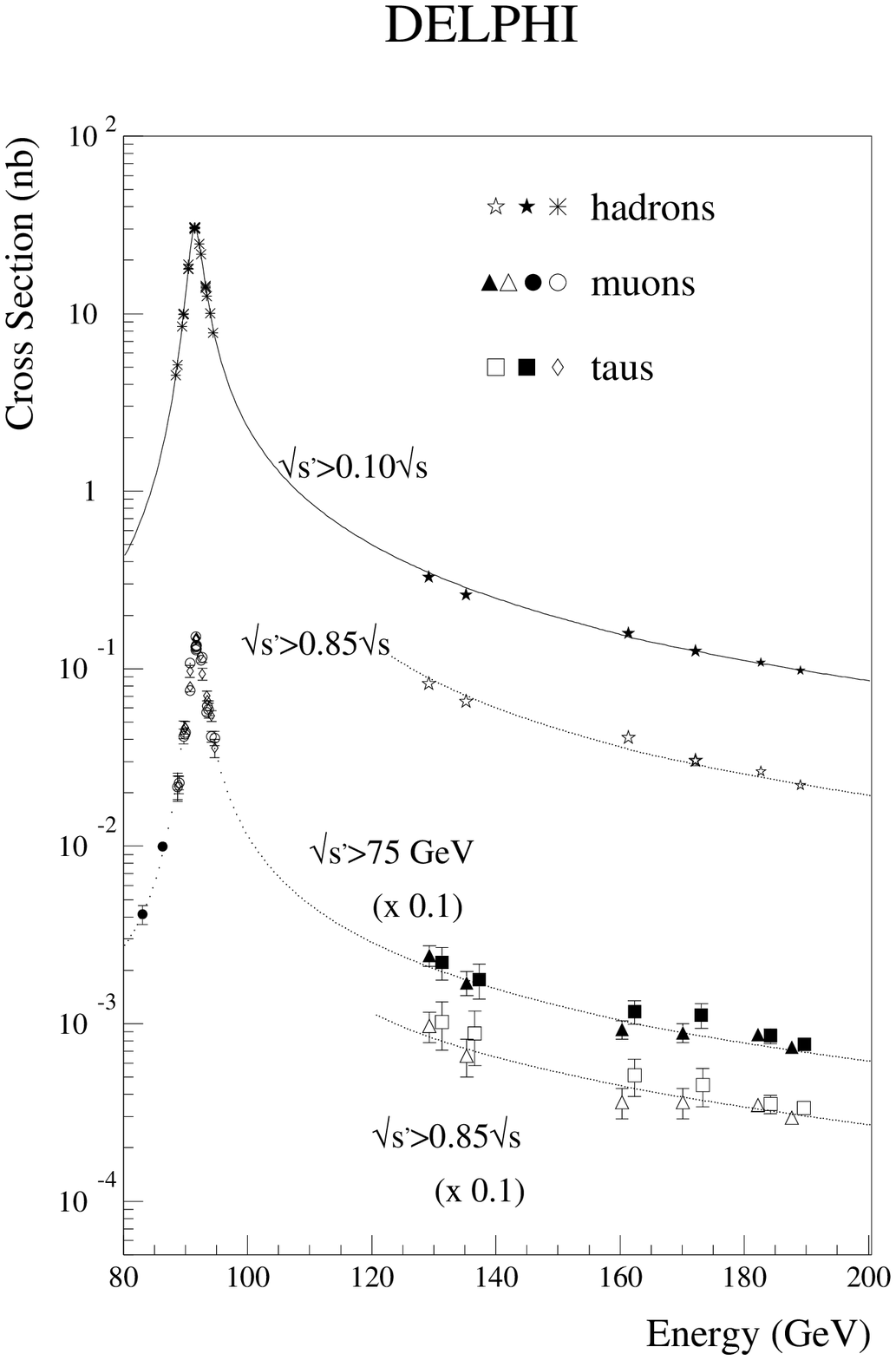}}
\end{center}
\caption{Total cross sections for fermion pair production in $\ee$ 
annihilation at LEP energies, from [120].}
\label{fig:tcross}
\end{figure}

I will discuss the three cases in turn, from large to small.  Consider
first the maxi-scale case.
  One might think that such large extra dimensions are
excluded by Cavendish experiments, but actually the best current limit is
only $R < 0.8$ mm ($M > 940 $ GeV) \cite{Mitro}.  More significant
constraints come from searches for quantum gravity effects at accelerators.
Two methods have been proposed.  The first is to search for 
processes in which a collision causes a graviton
to be radiated off the brane, carrying with it unobserved 
momentum \cite{GRW,MPP}. The 
simplest processes of this kind are
\beq
  \ee \to \gamma G \ , \qquad q \bar q \to g G \ ,
\eeq{missing}
where $G$ is a graviton, and these can be observed as missing energy processes
at $\ee$ and hadron colliders.  I will discuss the experimental status 
of this search in a moment.
The second is to search for a contact
interaction in fermion-fermion reactions due to graviton 
exchange \cite{GRW,Hewett,HLZ}.  The
coefficient of the induced contact interaction is 
model-dependent, to one cannot use this effect to set strict limits on $M$.
  But if the effect were there, it would be striking, 
 causing the cross sections in $\ee\to f \bar f$ to bend upward
as a function of energy, and also modifying the production angular 
distributions.  There is new data from LEP on the pair-production total
cross sections, but,  unfortunately for this purpose, it is in remarkable
agreement with the Standard Model prediction.  The new data from DELPHI
is shown in Figure~\ref{fig:tcross} \cite{DELPHItcross}.

%%%%%%%%%%%%%%%%%%%%%%%%%%%%%%%%%%%%%%%%%%%%%%%%%%%%%%%%%%%%%%%%%%%%%%%
%%%%%%%%%%%%%%%%%%%%%%%%%%%%%%%%%%%%%%%%%%%%%%%%%%%%%%%%%%%%%%%%%%%%%%%%%%%

On the other hand, 
 the cross sections for missing-energy processes can be
computed absolutely in terms of the gravity scale defined by \leqn{RMGrel}
and the number of extra dimensions $n$, so that bounds on these processes
allow us to place lower bounds on $M$.  In Table~\ref{tab:missing}, taken 
from \cite{CPP}, I give the best current limits on $M$ (at 95\% confidence)
from LEP and the Tevatron
and the sensitivity expected at LHC and at a 1 TeV linear collider.  The
LEP results correspond to new  limits announced at HEP99 \cite{Kounine}.
The first line of the table gives a set of bounds from an astrophysical 
source, the constraint that supernova 1987A did not radiate away most of its
energy in gravitons \cite{CP}.  This bound is very strong for $n=2$ but is
unimportant for larger $n$. I exclude cosmological bounds that are really 
constraints on the cosmological scenario.

%%%%%%%%%%%%%%%%%%%%%%%%%%%%%%%%%%%%%%%%%%%%%%%%%%%%%%%%%%%%%%%%%%%%%%%%%
\begin{table*}
\begin{center}
 \caption{Current and future sensitivities to large extra dimensions from 
 missing-energy experiments.  All values for colliders are 
expressed as 95\% confidence limits on the size of 
 extra dimensions $R$ (in cm) and the effective Planck scale $M$ (in GeV).
  For the analysis of SN1987A, we give probable-confidence limits.}
\bigskip
\begin{tabular}{l l  r  r  r }
\hline 
  Collider   & &  R / M  ($n=2$)  &  R / M  ($n=4$) &   R / M  ($n=6$) \\
\hline\hline 
   Present: &SN1987A   &  $3 \times 10^{-5}/50000$   & 
           $1 \times 10^{-9}/1000$ 
                 & $6 \times 10^{-11} /100  $ \\ 

 &LEP 2   &   $4.8\times 10^{-2}$ / 1200
                                       & $1.9\times 10^{-9}$ / 730 & 
                              $6.8 \times 10^{-12}$ / 530 \\ 
 & Tevatron  &   $5.5 \times 10^{-2}$ / 1140  & $1.4 \times 10^{-9}$ / 860 
              & $4.1 \times 10^{-12}$ / 780 \\
\hline
  Future: 
&  LC & $1.2\times 10^{-3}$ / 7700  & $1.2\times 10^{-10}$ / 4500 & 
                              $6.5 \times 10^{-13}$ / 3100 \\ 
 & LHC &  $4.5 \times 10^{-4}$ /12500  & $5.6\times 10^{-11}$ / 7500 & 
                              $2.7 \times 10^{-13}$ / 6000 \\ 
\hline
\end{tabular}
\end{center}
\label{tab:missing}
\end{table*}
%%%%%%%%%%%%%%%%%%%%%%%%%%%%%%%%%%%%%%%%%%%%%%%%%%%%%%%%%%%%%%%%%%%%%%%%%%%

In Table~\ref{tab:missing},
the sensitivity to missing-energy processes expected at the LHC is quite
remarkable.  These values cannot be completely trusted, for the usual reason
that the LHC cross sections integrate over very large momentum transfer
processes.  However, it is argued in \cite{CPP} that the values in the Table
are most likely to underestimate the LHC sensitivity.
 It was the original idea of Arkani-Hamed, Dimopoulos, and Dvali
that the size of the weak interaction scale should be set by the scale of 
$M$.  The LHC search for missing energy processes should provide a
sensitive test of this hypothesis.

I should note that the maxi- case throws away the grand unification scale
and all of the physics associated with it.  This includes the 
unification of coupling constants through the renormalization group.  
It also includes the suppression of neutrino masses and proton decay matrix
elements by the factor $\mw/M_\GUT$, a factor that arises naturally in 
the standard picture from the fact that these effects are mediated by 
dimension 5 operators.  New suppression mechanisms are needed if we have 
large extra dimensions.   Actually, this may be less a problem than an 
opportunity to discover new physical mechanisms; see \cite{ASchm} for an 
example.

\begin{figure}
\begin{center}
\leavevmode
{\epsfxsize=4.00truein \epsfbox{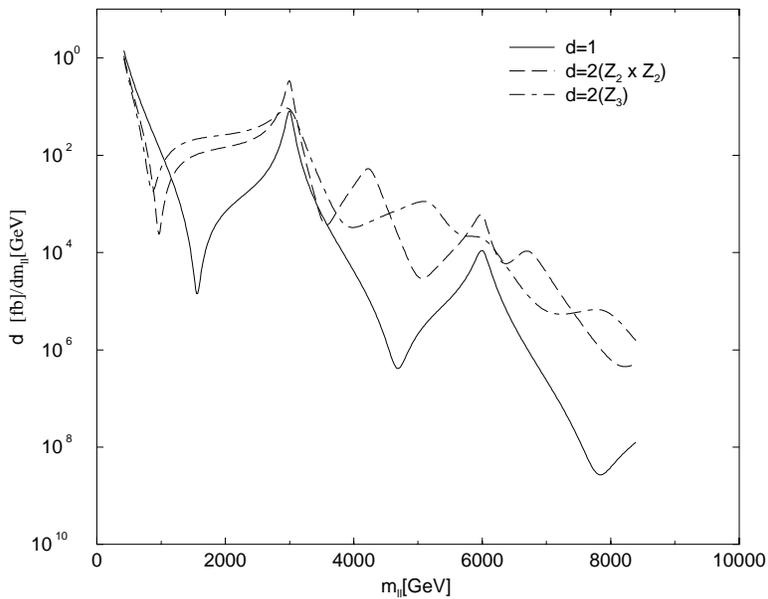}}
\end{center}
\caption{Dilepton mass spectra expected at the LHC due to Kaluza-Klein 
recurrences of the photon and $Z^0$, for $1/R = 3$ TeV, from [125].
  The solid curve is computed for 
one compactified dimension, the dashed curves for two cases of two 
compactified dimensions.}
\label{fig:Anto}
\end{figure}

We turn next to the midi- case.  In this class of models, the Standard Model
fields can consistently explore the extra dimensions.  Direct quantum 
gravity effects are inaccessible, but we should expect to see the excitation
of states of the photon, $Z^0$, and gluon with nonzero momentum in the 
extra dimensions.  These states appear in experiment as massive vector 
resonances, called `Kaluza-Klein recurrences'.  If the extra dimensions are
flat and have periodicity $2\pi R$, the masses of the these states are
$|\vec N|/R$, where $\vec N$ is a vector with integer components.  The 
spectrum of Kaluza-Klein recurrences is a Fourier transform of the shape of the
extra dimensions \cite{PYY,ABQ}.  Figure~\ref{fig:Anto}  shows the 
effect of these recurrences in producing resonances in the dilepton 
invariant mass distribution that would be observed at the LHC.

%%%%%%%%%%%%%%%%%%%%%%%%%%%%%%%%%%%%%%%%%%%%%%%%%%%%%%%%%%%%%%%%%%%%%%%
%%%%%%%%%%%%%%%%%%%%%%%%%%%%%%%%%%%%%%%%%%%%%%%%%%%%%%%%%%%%%%%%%%%%%%%%%%%

Finally, we come to the mini- case.  In this class of models, the direct
effects of the extra dimensions occur only at the grand unification scale
and cannot be observed experimentally.  A test of the hypothesis would have
to be based on a characteristic set of Lagrangian
parameters following from the geometry.  These parameters would provide a 
boundary condition for the renormalization group equations, and we would 
compare the results of integrating those equation to the weak interaction
scale with the results of our experiments.

\begin{figure}
\begin{center}
\leavevmode
{\epsfxsize=2.00truein \epsfbox{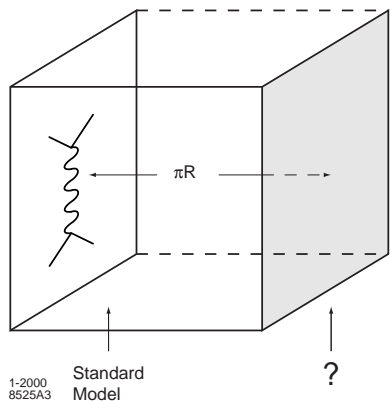}}
\end{center}
\caption{Picture of the universe viewed in the small following the 
ideas of Ho\v rava and Witten.  There is an extended 
fifth dimension, with  quarks, leptons,
and gauge fields bound to a wall on one boundary of the space.}
\label{fig:HW}
\end{figure}

Here is a concrete picture of how such a comparison could be made.  In the 
original model of Ho\v rava and Witten \cite{HW}, the geometry of Nature is 
effectively five-dimensional and has the form shown in Figure~\ref{fig:HW}.
The fifth dimension is bounded, and gauge bosons, fermions, and scalars are
bound to four-dimensional walls at the boundary of the space.  On one wall, we
would have the supersymmetric Standard Model.  What is on the other wall?
 Ho\v rava and Witten \cite{HW} proposed that this would be a natural place
to put the hidden sector responsible for supersymmetry breaking.

%%%%%%%%%%%%%%%%%%%%%%%%%%%%%%%%%%%%%%%%%%%%%%%%%%%%%%%%%%%%%%%%%%%%%%%
%%%%%%%%%%%%%%%%%%%%%%%%%%%%%%%%%%%%%%%%%%%%%%%%%%%%%%%%%%%%%%%%%%%%%%%%%%%

We must now ask, what supersymmetry spectrum follows from this hypothesis?
Two answers have been given in the literature.  Ho\v rava \cite{Horava}
has argued that
one should find the spectrum of gravity mediation in the no-scale limit.
However, this result has been criticized by Nilles, Olechowski, and Yamaguchi
\cite{NOY}, who have found large contributions to $m_0$ in his picture.
Randall and Sundrum \cite{RSanom} have argued that one should find the 
spectrum of anomaly mediation.  However, we have already seen that that
spectrum is not self-consistent and requires correction to produce positive
slepton masses.  Despite these problems with the answers that have been 
proposed up to now,  I believe that the question I have asked has a definite 
answer, and many theorists are now working to find it.  If someone
succeeds, the result will give a remarkable and concrete goal for the 
experimental studies of supersymmetry spectroscopy that we are soon to 
undertake.

\section{Lutheran sermon}

Tampere has a number of beautiful stone churches, and, in touring them, we
learned that sermons play an important role in the Finnish Lutheran traditions.
So I will conclude with a sermon.

For me, the most memorable part of the HEP99 meeting was a formal 
ceremony conducted by four young physicists representing the four LEP 
collaborations---Fabio Cerutti, Magali Gr\"uwe, Simonetta Gentile, and 
Mario Pimenta.  The title of the ceremony was: `Any sign of New Physics
in the 1999 LEP data?'.   The speakers were thorough, precise, and extremely 
well-informed. The answer to the question in the title was, no.

It is wrong to be cynical about such an exercise, but it is correct to be 
disappointed.  These speakers stood at the apex of a huge superstructure, 
representing more than a billion dollars of investment in equipment and
training, all focused on the goal of breaking through to the next layer of 
physics beneath the strong, weak, and electromagnetic interactions.
This time, we did not succeed.  What moral should we draw from this?

Most of the people in my audience for this lecture were still in grammar 
school in the 1970's. This was  a very different era in high energy physics,
with surprising discoveries and puzzles coming from experiment, forming a 
cloudy picture in which one struggled to the see the final outcome.
I was a graduate student in that period, and the excitement drew me in, away
from a perhaps more sensible career in the  physics of materials.  We do not
feel this sort of excitement in high-energy physics today, and many people 
now ask if it will ever return.

On the other hand, it is important to recognize that the experimental 
progress we have made in the 1990's is remarkable in another way.
It was often said in the early 1970's
 that the experimental picture was necessarily
unclear, because we were exploring a realm very remote from human experience.
Here I do not refer to the requirement that high energy phenomena need to be 
observed by complex detectors, but to the conceptual problem of visualizing
the basic objects that were used to construct theories---quarks, gluons,
heavy bosons, and the like.  It was thought that we could view these objects
only indirectly, by matching experimental results to abstruse theoretical
predictions.

In the 1990's we learned that this attitude is hogwash.  Quarks and leptons
may be unimaginably small, but with the right experiments, we can reveal all 
of the fine details of their behavior.   Especially through
 the program of precision 
experiments at the $Z^0$ resonance, we have been able to examine
the quarks or  leptons of each individual species,
shake their hands, and watch their dances.  However remote this
microworld is, we understand it pictorially, and with certainty.  It is 
important to add that  the means by which we have
achieved this understanding is that of using accelerators to go to 
the basic scale on which these objects act, and then looking and seeing  
what is manifest there.

In the process, we have learned that physics at this microscopic scale has 
a basis as rational as chemistry.  Quarks and 
leptons move, not by magic, but because there is a mechanism at work.
Experimentally, we have found the moving parts and exhibited their properties.
This understanding is very encouraging for the major unsolved aspects of the
behavior of the Standard Model, the breaking of electroweak gauge symmetry.
This phenomenon must also have a cause, and our experience in the 1990's 
tells us that, if we can patiently  continue our
investigations to higher energy, we can find it out.

The idea that there are mechanisms and reasons for physical phenomena, and 
that we can find the next one by searching to smaller distances, is an
article of faith.  As our experimental devices become more complex and
expensive, and as the time required to realize them stretches out, it 
becomes harder and harder to keep the faith.  The public wants results
on the nightly news.  Our allies in government work on the time scale of 
an election cycle; our colleagues in industry measure progress in 
`Internet time'.   It requires continuing effort to persuade them
that, though our enterprise moves much more slowly, it is in motion, and 
 toward 
important  goals.  But our hardest struggle is with ourselves and our
community, to press on to the next great period of discovery which is still
over the horizon.

In talking to many people in the experimental community, I sense a pessimism,
not about whether there is a next scale of physics, but about what we will
find there.  Just the Standard Model, they say, just the Higgs boson, just
the familiar
pattern that some theorist has set out.  The excitement of the 1970's has 
receded very far into the past, so far that it is difficult to imagine
that it will come again.

This is the reason that I have given so much attention in this lecture to the 
possibility of new space dimensions.  Though I have tried to motivate this
idea, I think that its major importance comes not so much because it must be
true as because it gives an example of how much we could have to learn, and 
how profoundly different the deep structure of the universe could be from 
what we now conceive.  There could really be unguessed secrets in the laws
of Nature.  And, these secrets are not hiding in the cosmos or on the large
scales of the universe, and not in rare materials or the organization of 
matter, but only at very small distances.  

This is where our accelerators will take us, if we can marshall our resources
and our intellectual strength.  Above all, we have to keep our belief
in our joint enterprise,  the belief
 that Nature has more wonders,  beyond our imagination,
which wait patiently for our tools to reach them.

\bigskip
I am grateful to Profs. Wolfgang Kummer and Matts Roos
and to the members of the HEP99 organizing
committee for giving me 
the opportunity to present this lecture, and to colleagues
too numerous to mention, both at SLAC and at HEP99, who have helped me to 
understand the topics discussed here.   I thank particularly Adam Falk,
Yuval Grossmann, and Zoltan Ligeti for discussions of CP violation.
My work is supported by the 
US Department of Energy under contract  DE--AC03--76SF00515.

\end{document}